| Title | From Electroculture to Plasma Agriculture: A Three-Century Arc Bridging Bertholon's Legacy with Contemporary Farming Advances |
|---|---|
| Authors | T. Dufour[1] |
| Affiliations | [1] *Laboratoire de Physique des Plasmas, 4 Place Jussieu, 75005 Paris, Sorbonne Université, CNRS, Ecole polytechnique.* |
| Correspondence | thierry.dufour@sorbonne-universite.fr |
| Ref. | Compte-Rendus Mécanique de l'Académie des Sciences, Volume 354 (2026), pp. 89-116 |
| DOI | https://doi.org/10.5802/crmeca.331 |
| Summary | This review traces the historical trajectory of electricity in agriculture, from the earliest observations of electrical phenomena to the emergence of cold plasmas. Looking back to Antiquity and then to the Enlightenment, it underlines Abbé Bertholon's 18th-century efforts to channel atmospheric electricity to stimulate crops, using devices such as the electro-végétomètre. Although these early electroculture experiments relied on neither quantitative dosimetry nor rigorous methodology, they foreshadowed the idea of a controlled transfer of electrical energy to plants. Then the review examines the historical development of galvanism, electrochemistry, and the physics of gaseous discharges throughout the 19th and 20th centuries, which collectively laid the foundations for contemporary cold-plasma technologies. In the 21st century, plasma agriculture has emerged as an interdisciplinary approach integrating electrical, chemical, radiative, thermal, and fluid-mechanical effects. Applications include seed treatment (preconditioning, seed priming), stimulation of plant growth, soil and water treatment, and decontamination of agri-food products. The review thus reassesses Abbé Bertholon's contributions as those of a methodological precursor and shows how his intuition of a "vivifying electricity" resonates with modern cold-plasma science. Finally, it argues that plasma agriculture can transform an Enlightenment intuition into a reproducible experimental framework for sustainable agriculture and food safety. |
| Keywords | Electroculture; Plasma agriculture; Cold atmospheric plasma (CAP); Abbé Bertholon; Dose–response relationships; Agricultural innovation; Plasma-activated water (PAW); Reactive oxygen and nitrogen species (RONS); Seed priming; Sustainable agriculture |

# 1. Electricity and agronomy: a historical trajectory from Antiquity to the 21st century

## 1.1. Electricity (Antiquity-18th century): from experimental spectacle to metrological instrumentation

From Antiquity to the end of the 18th century, the study of electrical phenomena evolved from a qualitative curiosity to a quantitative, instrument-based practice. Early authors such as Theophrastus (*De lapidibus*) and Pliny the Elder (*Natural History*) noted the attraction of straw and other light objects by rubbed amber (*elektron*) and jet (gagates), while medical authors (e.g., Scribonius Largus, Galen) reported the analgesic use of electric fish (torpedoes) **[1–3]**. Medieval lapidaries and encyclopedias widely transmitted these observations in descriptive form, with Albertus Magnus' *De mineralibus* being the emblematic example **[4]**. During the Renaissance, Girolamo Cardano (*De subtilitate*, 1550) and Giambattista della Porta (*Magia naturalis*, 1558/1589) expanded the inventory of "electric bodies" (amber, jet, diamond, sapphire) capable of attracting small particles after friction **[5, 6]**. The thread of observation thus remained continuous until the end of the 16th century but was essentially qualitative: there were no calibrated instruments and procedures were poorly standardized.

A conceptual consolidation began with William Gilbert's work *De Magnete* (1600), who coined the term "electricus" and clearly distinguished electrical attractions from magnetism **[7]**. In the 17th and early 18th centuries, friction machines multiplied demonstrations and made the production of charges more reliable and visually convincing: Otto von Guericke's sulfur globe (1660s) and Francis Hauksbee's luminous glass globes (1705–1710) are canonical examples **[8, 9]**. Stephen Gray's experiments (1729) established conduction and insulation as distinct behaviors and Charles-François de Cisternay du Fay (1733) organized the effects under a "vitreous/resinous" duality **[10, 11]**. Nevertheless, according to subsequent metrological standards and in the absence of sustainable charge storage and common measurement conventions, electricity remained a theater of demonstration (sparks, "electric kiss", bristling hair).

It was only in the 1740s that advances in instrumentation created conditions for cumulative experiments and comparative methods. The Leyden jar, developed independently by E.-G. von Kleist and P. van Musschenbroek (1745–1746), was the first reliable device for storing electrical charge, making it possible to repeat protocols and compare outcomes systematically (Figure 1a). Linking several jars in parallel to form a "battery" further increased storage capacity and enabled more powerful discharges (Figure 1b) **[12–14]**. At the same time, electrostatic machines became both more standardized and more powerful, exemplified by Nairne's cylinder (1773; Figure 1c) and van Marum's monumental generator (1784). These devices reflected not only the growing technical capacity of the sources but also a nascent ambition to measure and quantify electrical effects **[15, 16]**. Public demonstrations provided a bridge between spectacle and experiment: on March 14, 1746, Abbé Nollet famously discharged a Leyden jar through a chain of the king's guards at Versailles and then through Carthusian monks. Such iconic moments dramatized electricity as a visible, collective phenomenon while simultaneously consolidating experimental techniques and a shared community of practice [17, 18].

In parallel, metrological tools multiplied and gradually shifted attention from spectacle to measurement. Franklin's demonstration of the equivalence between lightning and electricity and his promotion of the lightning rod (1750–1752), redirected interest toward the controlled delivery of charge. These claims also sparked a widely read debate with Nollet [17, 21, 22]. Building on this shift toward quantification, Lane's spark-reading electrometer (1767) allowed the quantification of Leyden jar discharge, while the Henley quadrant (1770) introduced a single-index measure of repulsion that was rapidly disseminated by Nairne and documented by Priestley (1772) [23–25]. H.-B. de Saussure extended electrometers to atmospheric studies during the 1760s–1780s [26]. This quantitative turn culminated with Coulomb's torsion balance (1785), which yielded the first precise law of electrical forces [27]. At the same time, synthesis works such as Priestley's *History and Present State of Electricity* (1767/1769), Cavallo's *Complete Treatise of Electricity* (1777) and Beccaria's *Dell'elettricismo artificiale e naturale* (1753) organized the expanding literature and, crucially, framed "artificial" and atmospheric electricity within a single analytic perspective [14, 28–30]. Collectively, these developments established a proto-metrology of electricity: systematic measurement practices and conceptual unification were now in place, even if standardized units and calibrated scales were only beginning to emerge.

Having established tools for storage, delivery and measurement, the first applications to living systems began to emerge. In 1783, Bertholon's *De l'électricité des végétaux* introduced the "electro-végétomètre," a device designed to capture and channel atmospheric electricity toward crop beds, as detailed in Section 2.3 [31]. In his treaty, Bertholon mentions a broader set of contemporary experiments that suggested electricity had a stimulating effect on vegetation. Stephen Demainbray in the 1740s electrified myrtles and observed accelerated growth, results confirmed by Abbé Nollet on mustard seeds and Jean Jallabert on hyacinth and daffodil bulbs. Georg Matthias Bose and Abbé Menon described similar effects on shrubs and buds. Later, Edouard-François Nunebert compared non-electrified and electrified onion bulbs, noting faster germination and taller stems, while Abbé d'Ormoy reported thicker bases and stronger roots across several species. Jean-Paul Marat claimed positive results on cress and lettuce and Franz Karl Achard made similar observations on crop plants [31].

By the close of the 18th century, a rudimentary metrology of electricity had thus taken shape: storage through the Leyden jar, delivery via standardized machines and conductors and measurement with electrometers and the torsion balance. This transition, from spectacular display to instrumented practice, prepared the ground for the galvanic and electrochemical breakthroughs of the 1790s (Section 1.2) and, more broadly, for later process-oriented uses where measurable inputs and controlled exposures would become central.

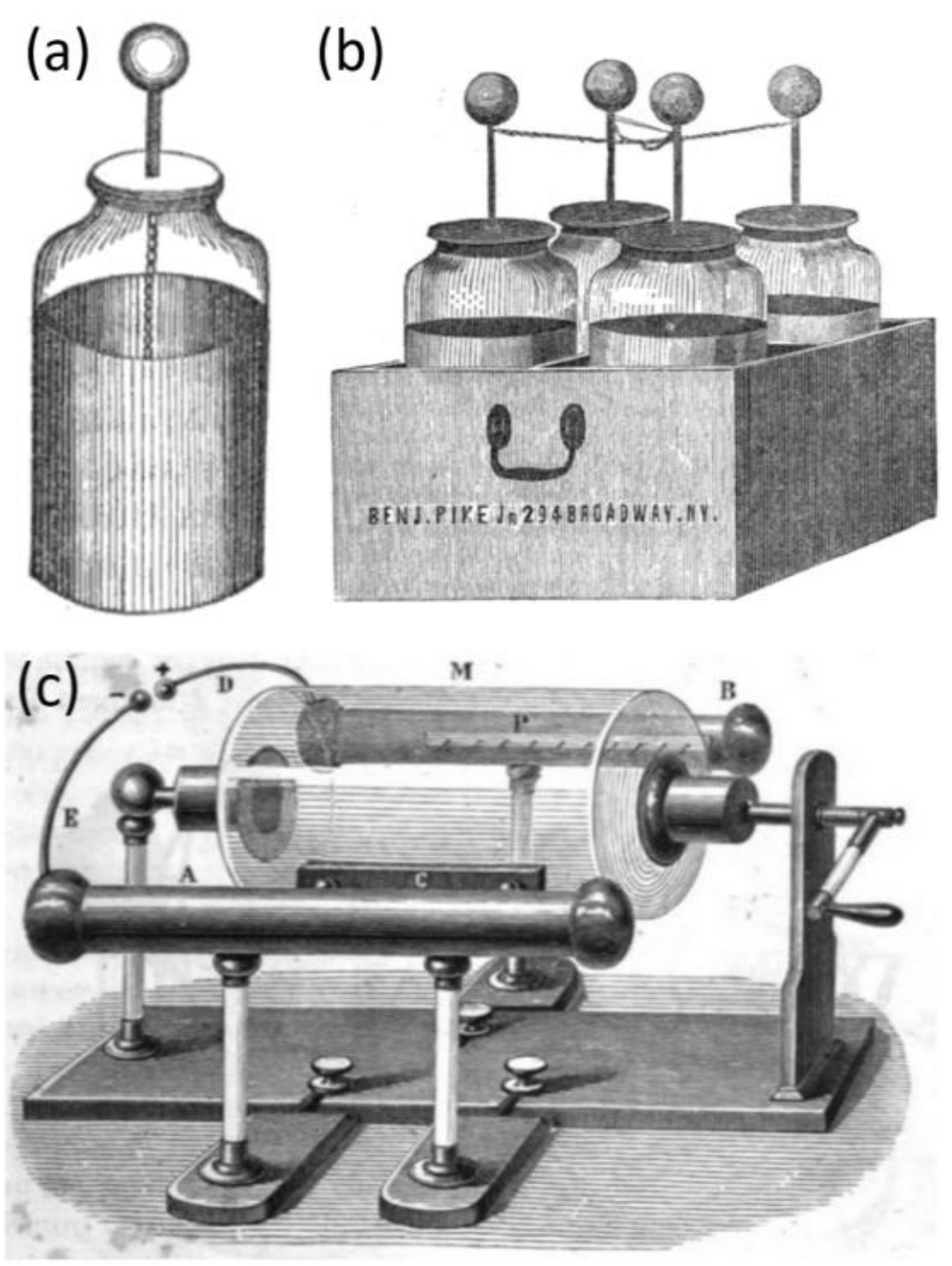

***Figure 1. (a) Leyden jar: a glass vessel acting as the dielectric, with an inner tin-foil lining connected by a chain to the top terminal; an outer foil (grounded, not shown here) forms the second armature. Charge is stored between the two foils separated by the glass [19]. (b) Battery of Leyden jars ("Franklin battery") in a wooden case. The inner electrodes of the jar are bus-bar connected; the outer foils are commoned through the box, forming a parallel array that increases total capacitance and stored energy for stronger discharges [19]. Nairne's cylinder electrical machine (c. 1770s). Hand-cranked glass cylinder rubbed by a cushioned exciter generates charge; collecting combs feed the insulated prime conductors (foreground bars). Polarity can be selected by connecting either end of the cylinder and a Leyden-jar terminal allows storage and discharge for demonstrations and experiments [20].***

## 1.2. 1790-2000: From galvanism to controlled discharges and their processes

Following the debates at the end of the 18th-century debates, Volta's invention of the pile (1799–1800; Figure 2a) marked a decisive turning point. For the first time, experimenters had access to a continuous, controllable and reproducible source of current [32]. Through the 19th century, this foundation was reinforced by the standardization of instruments and units: precision galvanometers and bridges, standard cells and the practical volt–ohm–ampere system together provided a calibrated metrological framework. Electricity, once confined to episodic demonstrations, was transformed into a measurable flow, available on demand and suitable for serial, comparable protocols.

On this basis, Faraday formalized the laws of electrolysis in the 1830s, establishing both a conceptual and an operational framework for the conversion of electricity into chemistry [36]. In biology, the same stability of supply opened up new paths. In 1873, Burdon-Sanderson recorded action potentials in Dionaea muscipula, showing that plant movements could be associated with discrete electrical signals [37]. A generation later, Bose extended this program into plant electrophysiology, developing instruments and methods to detect, transmit and analyze plant signals (Figure 2b) [33]. The idea that an electrical stimulus could modulate measurable plant functions (movements, conduction, responses to stress) was thereby firmly established, laying the groundwork for systematic bioelectrical experimentation.

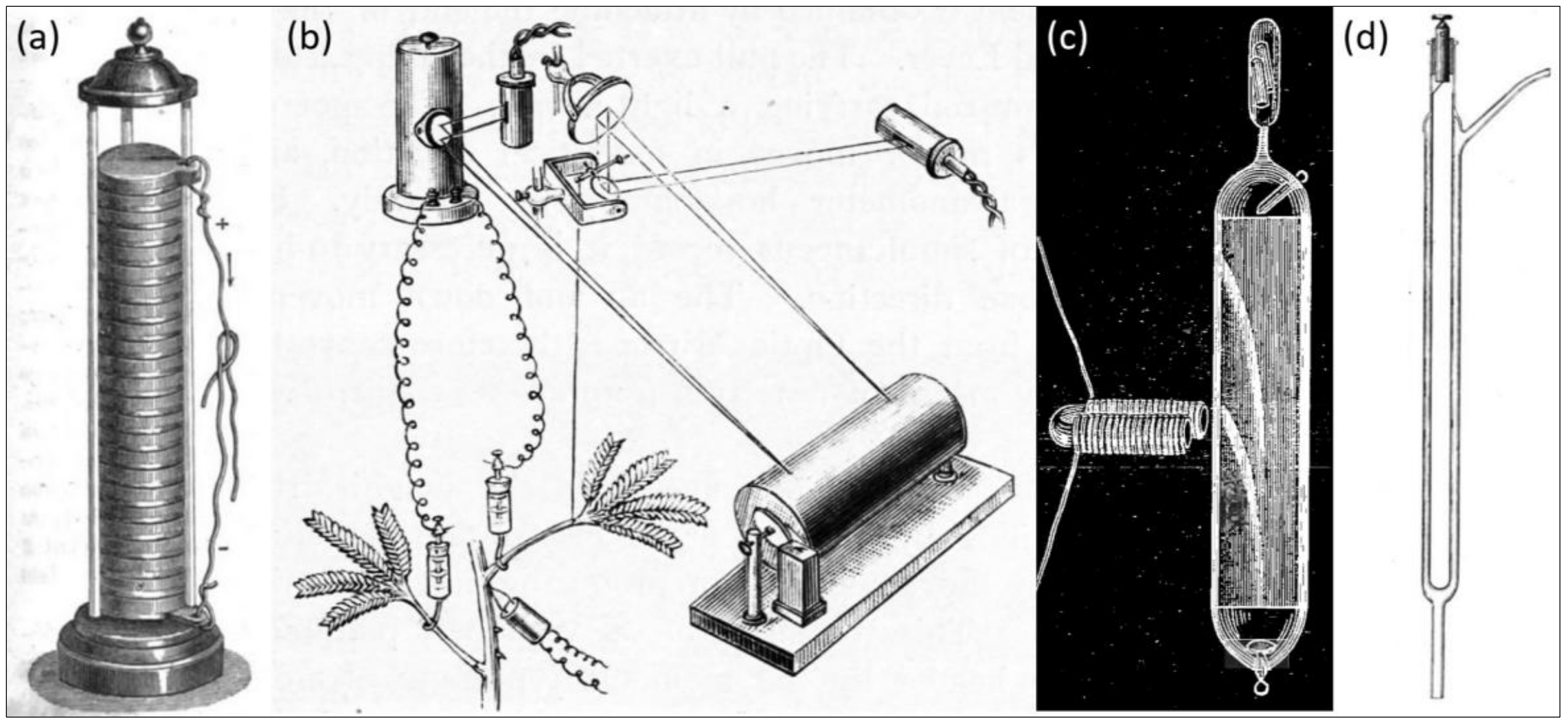

*Figure 2. (a) Voltaic pile (1800): column of alternating zinc-copper disks separated by brine-soaked washers. Polarity is set by couple orientation: the end toward the zincs is negative, toward the coppers positive (modern convention). Removing terminal plates does not change polarity. Side leads provide (+)/(-) connections [20]. (b) Bose's apparatus for simultaneous mechanical and electrical responses: an induction-coil stimulator; a mirror galvanometer records the electrical trace, an optic levers the movement; lamp-reflected beams write on a rotating kymograph; latency yields stimulus-response relations [33]. (c) Crookes tube (high vacuum): a phosphorescent screen and a slotted mica diaphragm form a beam of cathode rays; a magnet deflects the beam, demonstrating negative charge and magnetic sensitivity [34]. (d) Werner von Siemens's first ozonizer (1857): glass discharge tube operating a "silent" dielectric-barrier discharge; dry air/oxygen traverses the annular gap where ozone is produced [35].*

From the second half of the 19th century, a true metrology of discharges began to emerge, extending far beyond the simple spark. In 1889, Paschen formulated his law relating breakdown voltage to pressure and electrode spacing, giving experimenters a predictive tool for designing and scaling apparatus [38]. A decade later, Townsend's ionization theory (1897–1901) clarified the avalanche multiplication of electrons and established the different regimes of gaseous discharge [39, 40]. In parallel, new instruments opened wider experimental spaces: Geissler's tubes and Crookes's rarefied-gas experiments (Figure 2c) revealed the diversity of luminous and conductive phenomena under reduced pressure [34, 41]. As early as 1857, Werner von Siemens described the dielectric barrier discharge (DBD) also known as "silent discharge", as a practical source for ozone generation (Figure 2d) [35]. By 1928, Langmuir gave this domain its modern vocabulary, introducing the term plasma to describe a quasi-neutral ionized gas [42]. Taken together, these advances transformed the conception of electrical discharge. What had once been understood as a fleeting spark now appeared as a controllable medium, with stabilizable regimes and tunable energy input.

On this metrological foundation, structured chemical applications quickly followed. Siemens's dielectric barrier discharge ozonizers, developed in the late 19th century, enabled large-scale water ozonation and gave the "silent discharge" its first industrial role [43, 44]. A few years later, the Birkeland–Eyde arc (1903, Figure 3a) harnessed high-temperature plasmas to oxidize atmospheric nitrogen, producing nitrates and nitrous acids as precursors for fertilizers [45].

Alongside these advances, low-pressure mercury-vapor lamps systematized the germicidal use of ultraviolet radiation: their characteristic 254 nm emission was applied to air and water disinfection in experimental setups such as Rentschler's (Figure 3b) [46–48].

At the beginning of the 20th century, electroculture experienced renewed interest, with attempts to exploit atmospheric electricity as a stimulus for plant growth. Devices ranged from Lemström's overhead networks of charged wires above fields to Christofleau's so-called "atmospheric" antennas [49, 50]. As shown in Figure 4a, the latter were designed to channel "natural electricity" from diverse sources (wind, clouds, sunlight, rain and even telluric currents) into the soil. In practice, however, the antenna functioned primarily as a charged conductor exposed to atmospheric potential gradients, behaving more like a lightning rod than a genuine energy transfer system. Though scientifically unfounded, such instruments helped popularize the notion of directing "sky electricity" toward crops. Figure 4b depicts their application to grapevines, while Figure 4c illustrates reports of enhanced bean growth compared with controls fertilized only with manure. Yet these outcomes must be interpreted with caution: fertilization regimes, soil heterogeneity, seasonal effects, rainfall and cultivation practices were poorly controlled and device geometries and polarities varied considerably. Consequently, these trials are best understood as part of a genealogy of ideas and technical imagination, rather than as rigorous demonstrations of efficacy. Their true legacy lies in shaping the concept of a controlled transfer "sky → soil/plant" and in fostering agronomic observation (emergence, yields), foreshadowing later, quantitatively controlled approaches based on explicit, traceable dose metrics.

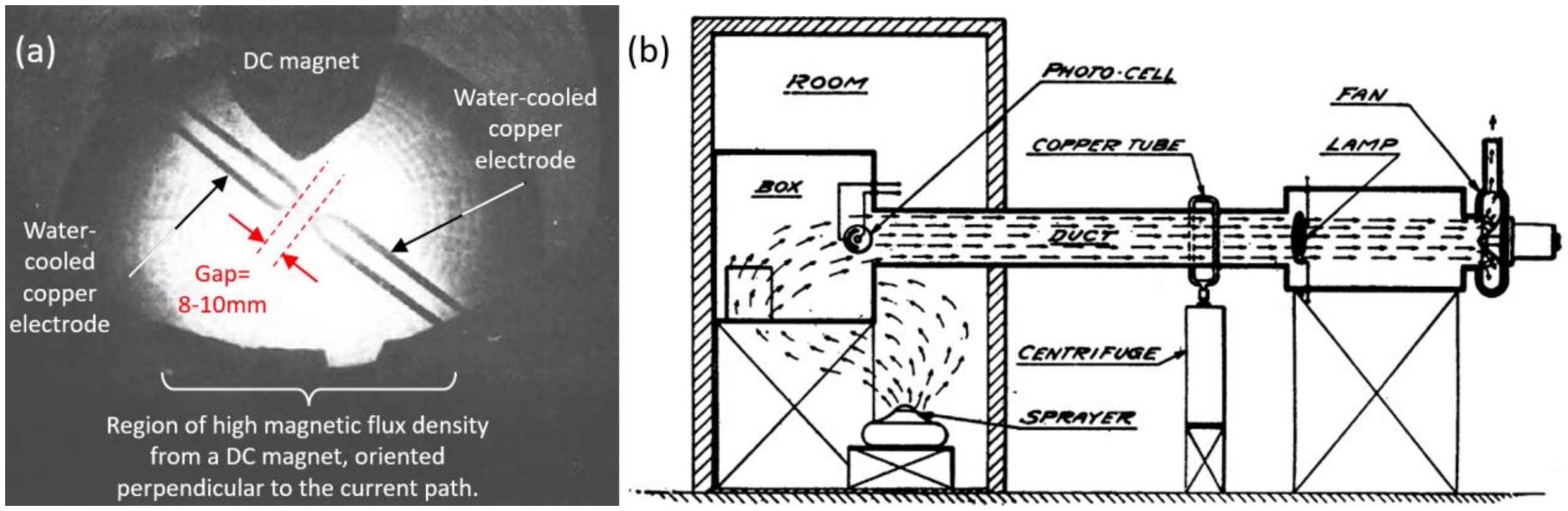

***Figure 3. (a) Birkeland–Eyde disc flame. An AC arc between water-cooled copper tips is blown into a rotating circular sheet by a transverse magnetic field between the pole faces. The time-averaged luminous disc is the hot reaction zone used to form NO in air [45]. (b) UV-germicidal aerosol test duct (15 ft × 1 ft): room air seeded with atomized E. coli passes a mixing box and a low-pressure mercury lamp (UV-C ≈254 nm); a perforated copper traverse tube feeds Wells air centrifuges; a tantalum photocell monitors irradiance; an axial fan sets the flow [48].***

This historical current of electroculture, however, did not directly evolve into or compete with plasma-based technologies. The latter emerged independently much later, at the end of the 20th century, driven not by the legacy of "sky electricity" but by advances in plasma physics and its biomedical applications. Whereas electroculture remained speculative and pre-metrical, cold plasma research developed within a rigorous framework of controlled discharges and quantitative dosimetry. Hence, from the 1970s onward, cold plasmas began to be investigated as low-temperature agents for sterilization and surface decontamination. Initial work focused on low-pressure plasmas to inactivate microorganisms on medical materials and devices. By the 1990s, these methods entered healthcare practice, most notably through hydrogen peroxide gas-plasma sterilizers for heat-sensitive instruments **[51, 52]**. In the mid-to-late 1990s, a further step was achieved with the emergence of atmospheric sources (including DBDs, glidarc reactors, plasma jets and RF discharges) that operated directly in open air. These systems demonstrated robust microbial inactivation and, crucially, allowed treatment conditions to be parameterized in terms of applied power, frequency, working gas and exposure time **[53]**. This trajectory, from biomedical sterilization tools to well-characterized atmospheric plasmas, laid the groundwork for extending plasma technologies into agronomy, giving rise to the modern field of plasma agriculture.

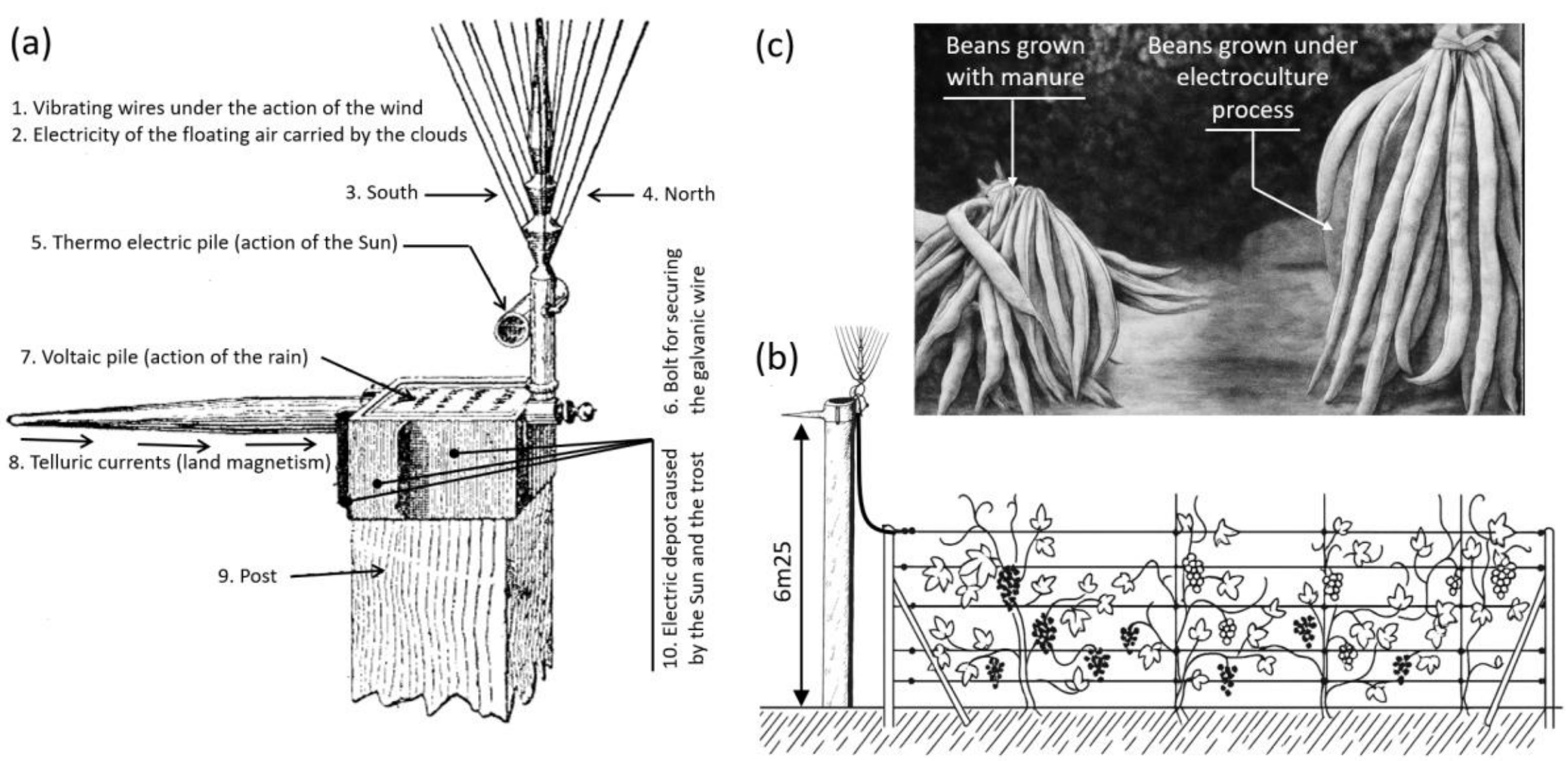

***Figure 4. (a) Schematics diagram introducing Christofleau's atmospheric antenna: a passive mast with pointed collector, galvanic wire and ground network, claiming to harness atmospheric, thermoelectric, rain-galvanic and telluric sources to deliver weak fields/currents to crops [50]. (b) Christofleau-style atmospheric antenna for vines: mast tipped with a multi-point collector feeds a down-lead to a trellised network of horizontal wires around grapevines, forming a passive aerial-ground circuit intended to deliver weak atmospheric fields/currents to the crop. Adapted from [50]. (c) Digitally-restored photograph (for clarity) of beans either grown with manure (left) or grown without manure but under Christofleau's atmospheric electroculture antenna (right). Adapted from [50].***

### 1.3. The 21st century: the rise of plasma agriculture in a physico-chemical framework

Plasma agriculture designates a contemporary research field that applies cold plasma sources (and their derivatives such as afterglows and plasma-activated water) to seeds, plants, soils, irrigation systems and agri-food chains. Unlike earlier traditions of electroculture or electrostatic spraying, which relied on static fields or charge deposition, plasma agriculture is explicitly grounded in plasma chemistry, physico-chemical dosimetry and dose–response control. Bibliometric analysis (Figure 5a) shows that the term “plasma agriculture” gained traction only after the mid-2010s, marking the consolidation of a distinct research community and vocabulary.

Cold plasmas are non-thermal discharges whose efficacy arises from the co-action of five properties (Figure 5b):
• Electrical: transient fields, ionization fronts and pulsed currents govern initiation and charge transfer
• Chemical: cold plasmas generate reactive oxygen and nitrogen species (RONS) both short-lived (OH, O, NO) and long-lived ($H_2O_2$, $O_3$, $NO_2^-$ /$NO_3^-$), with tunable lifetimes and fluxes
• Radiative: UV and VUV photons couple directly to DNA/protein targets or water interfaces
• Thermal: bulk gas remains near ambient, though local heating (10–50 °C above background) can be engineered or suppressed
• Fluid-mechanical: ionic wind and microjets enhance mass transfer at plasma–liquid and plasma–tissue boundaries.

Depending on the source (DBD, jet, corona, glidarc) and feed gas, these modalities combine to define the “plasma cocktail” at seed–water and plant–air interfaces [54–56].

By 2025, plasma agriculture is organized around four application axes:

- Seed treatment exploits plasma-induced surface modification (enhanced hydrophilicity, micro-etching) and low-dose RONS/UV to accelerate germination kinetics and reduce microbial load [57]. Direct dry-seed exposure, post-imbibition treatment and immersion in plasma-activated water (PAW) are distinguished approaches [58]. PAW chemistry is now routinely quantified by dose tracers ($H_2O_2$, $NO_2^-$/$NO_3^-$, pH, conductivity), providing reproducibility across laboratories [59].
- Plant treatment applies CAP, afterglows and PAW under strict thermal control to elicit signaling rather than damage. Contemporary work demonstrates redox priming that activates antioxidant and defense pathways and improves vigor in model and crop species (e.g., $Ca^{2+}$ transients in Arabidopsis with PAW; tomato seedlings under CAP/PAW) [60–62]. Electrical inputs (kV, kHz, W), flow, gap, exposure time and effluent temperature are systematically coupled to PAW tracers, in line with international plasma-agriculture roadmaps.
- Media treatment focuses on irrigation water, soil and air. CAP and PAW enable gentle disinfection, dose-controlled inputs and modulation of microbial communities. Recent syntheses map PAW physico-chemistry and antimicrobial action across matrices [63]; field-adjacent studies demonstrate impacts on irrigation water, soil properties and crop stress responses [64].
- Agri-food treatment uses CAP (DBD, plasma jets) and PAW to sanitize produce, packaging and equipment while extending shelf life without thermal stress. Efficiency scales with dose (voltage, frequency, gas, flow/gap) and exposure time, primarily through RONS and UV [65]. “In-package” plasma retains reactive species in sealed headspaces, boosting microbial inactivation with limited quality loss [66]. On open lines, plasma jets reduce biofilms on food-contact surfaces while PAW complements as a produce wash with minimal residues. Industrial uptake by 2025 is contingent on validated protocols, material compatibility and maintaining organoleptic quality [67].

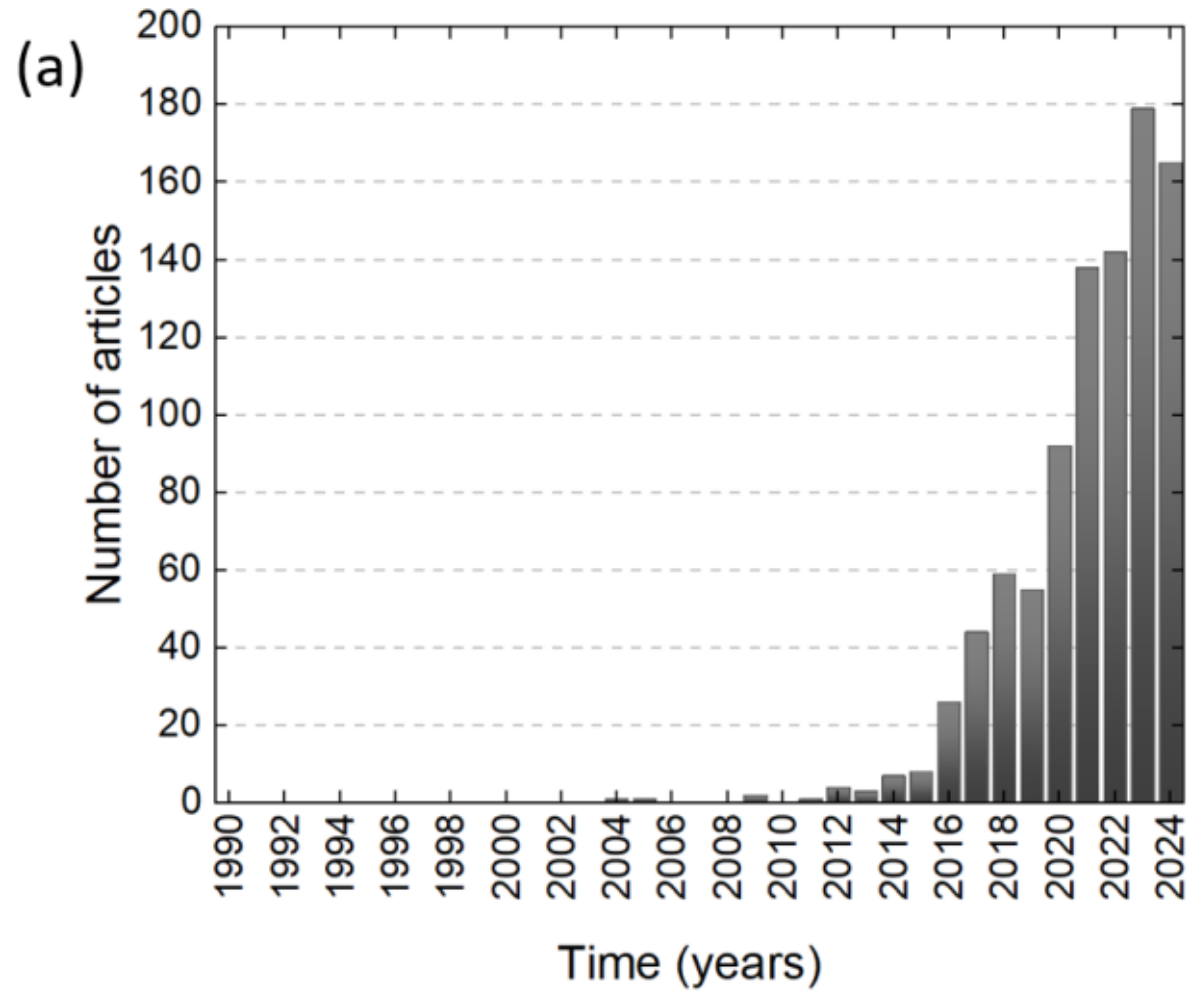

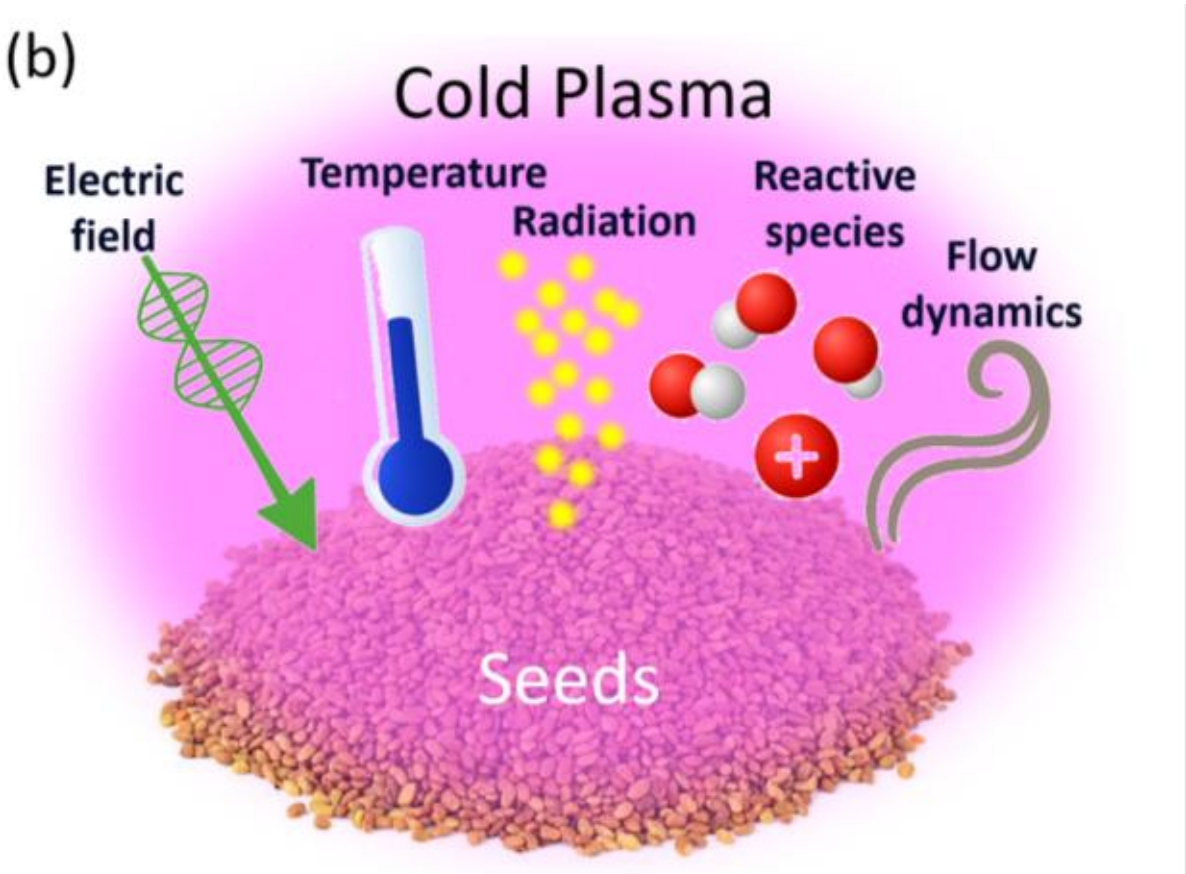

***Figure 5. (a) Annual number of publications (1990-2024) containing the exact expression “plasma agriculture,” as indexed by Google Scholar (Results retrieved on 18 August 2025, patents and citations excluded). (b) Schematic of a pile of seeds treated by cold atmospheric plasma (CAP) with its 5 coupled properties: electrical, chemical (RONS), radiative (UV), thermal (near-ambient) and fluid-mechanical.***

In contemporary plasma agriculture, progress is defined by a physics-based framework of explicit dosimetry: quantitative, traceable dose metrics (such as RONS concentrations in PAW ($mol \cdot L^{-1}$), UV fluence ($mJ \cdot cm^{-2}$), electrical or energy dose ($J \cdot cm^{-2}$) and exposure time (min)) are measured at the target. These are coupled with controlled source parameters (voltage, frequency, power, feed gas, flow, gap and effluent temperature) to establish reproducible dose–response relationships. This approach contrasts sharply with 18th-century electroculture, where no measurements of field strength, exposure time or chemical by-products were attempted. To appreciate the extent of this methodological shift, the following Section 2 revisits the early but pre-metrical program of Abbé Bertholon, one of the first to propose the systematic use of atmospheric electricity in agriculture.

# 2. Reappraising the Abbé Bertholon's contribution to agricultural electricity (1780s)

## 2.1. A brief scientific biography

Pierre-Nicolas Bertholon de Saint-Lazare (1741–1800), a priest, physicist and Enlightenment pedagogue, built his career in Montpellier (France), where he taught experimental physics at the chair of the États de Languedoc and within the Royal Society of Sciences **[68]**. During the 1780s he emerged as a central figure of experimental electricity in the city, notable for his focus on applications rather than spectacle. Appointed in 1782 to the Society's chair of physics, he combined public instruction with carefully staged demonstrations, framing his program around three domains: meteorological electricity, animal electricity and plant electricity, each approached with an emphasis on practical use **[69]**.

As part of this program, Bertholon designed the electro-végétomètre, a device intended to capture and distribute "atmospheric electricity" to plants. He also promoted "medical electricity", devising therapeutic practices such as positive and negative baths and even proposed a classificatory scheme for diseases like "electrical" or "non-electrical" **[31, 70]**. His writings, first published in 1780, were widely circulated in both French and translated editions, extending his influence beyond Montpellier.

Bertholon also participated actively in civic and technical debates. He advocated for the adoption of lightning rods in Languedoc and, by analogy, suggested networks of pointed rods as earthquake prevention devices, a proposal that drew attention as far as Germany (Wiedebourg) **[69]**. He engaged in the early aerostatic movement, launching a hot-air balloon with Jean-Antoine Chaptal in Montpellier on 23 November 1783 and testing parachutes with Lenormand from the Babotte Tower.

His intellectual interests extended further, encompassing geology, volcanism and regional natural history, reflecting the broad culture of inquiry encouraged by the Montpellier Society. In 1787, together with Boyer, he co-edited the popular science journal *La nature considérée sous ses différents aspects*, underscoring his role as a mediator between scientific institutions, laboratories and wider publics **[71]**.

## 2.2. Bertholon's thesis: electricity as a stimulus to vegetation

In the late 18th century, as Leyden jars, electrometers and electrostatic machines were establishing a first metrology of electricity (Section 1.1), Pierre-Nicolas Bertholon advanced a striking thesis: that atmospheric electricity was a constant reservoir of electricity capable of stimulating the entire physiology of plants. He argued that key functions (germination, sap circulation, nutrition, secretion, growth, flowering, fruiting and transpiration) could all be activated by this natural resource **[31]**. Moreover, artificial electricity generated by friction machines could, in his view, produce equivalent effects. Electricity was thus conceived as a subtle "fluid", often likened to fire, whose action might be harnessed oriented and applied through engineered devices. Bertholon's treatise was structured in three parts:

(1) To demonstrate the existence and influence of atmospheric electricity;
(2) To catalogue its effects on plant functions (including germination, sap flow, secretions, sensory qualities in the 18th-century sense and "negative electricity");
(3) To operationalize these effects through dedicated instruments, most notably the electro-végétomètre, designed to capture, conduct and distribute atmospheric electricity to crops.

To substantiate this program, Bertholon assembled three types of evidence. First, qualitative correlations: storms, rain, snow, hail and fog were thought to deliver electrical fluid to plants. He noted that "more electrical days" coincided with faster germination, though without controlling for cofactors such as temperature, humidity, light or rainfall. Second, instrumented observations: he employed electrometers, kites and electroscopes to register atmospheric charge and compared these signals with stages of plant development. He described precautions (stable weather, humidity, temperature) yet acknowledged variability, preferring broad "general results" to unstable numerical series. Third, mechanistic conjectures: electricity was said to drive sap circulation and secretion through an "electrical affinity" between molecules and vessels: an attempt, within the conceptual vocabulary of his time, to articulate a theory of plant interfaces.

From a historical perspective, the strength of Bertholon's work lies in its unifying vision. He sought to connect atmospheric phenomena with plant physiology and to design an agronomic technology of electricity. His vocabulary drew on engineering concepts such as insulation, pointed collectors, conductors and distribution networks and for his time this represented a genuine methodological effort. The limits were equally clear: no dosimetry of field strength, charge or energy was attempted, the roles of different agents (electric fields, ozone, UV) were not distinguished and much reasoning remained analogical rather than quantitative. Even Bertholon conceded that his "constant results" were better understood as tendencies than as conclusive data.

Yet the legacy is not negligible. By framing electricity in agronomic terms and articulating an operative vocabulary of capture, conduction and distribution, Bertholon anticipated the logic of controlled transfer that resonates with later traditions. In contemporary cold-plasma agriculture, electrical parameters, physico-chemical species and dose–response relations are explicitly integrated. Seen in this continuum, Bertholon's program was not merely speculative electroculture but an early, if limited, attempt to transform atmospheric electricity into a measurable agronomic process.

## 2.3. Bertholon's electroculture devices and operating procedures

The first electroculture devices to be technically described and illustrated appear in Bertholon's *De l'électricité des végétaux* (1783). Because the original explanations, written in 18th-century French, are often obscure or misleading for modern readers, this section offers a reformulation in contemporary technical language. The goal is not to modernize the science, but to clarify the intended mechanisms, stabilize terminology and highlight the operational logic of these instruments.

### 2.3.1. Dry-route devices: the electro-végétomètre & the orchard discharge method

Bertholon's flagship instrument was the electro-végétomètre: an apparatus which, despite its name, was not a measuring instrument: it recorded neither electric field, nor electrical charge, nor current. It was a passive collector–distributor of "atmospheric electricity", designed to capture, conduct and spread charge over crops, with no readout and no dosimetry. Bertholon designed two models of this device.

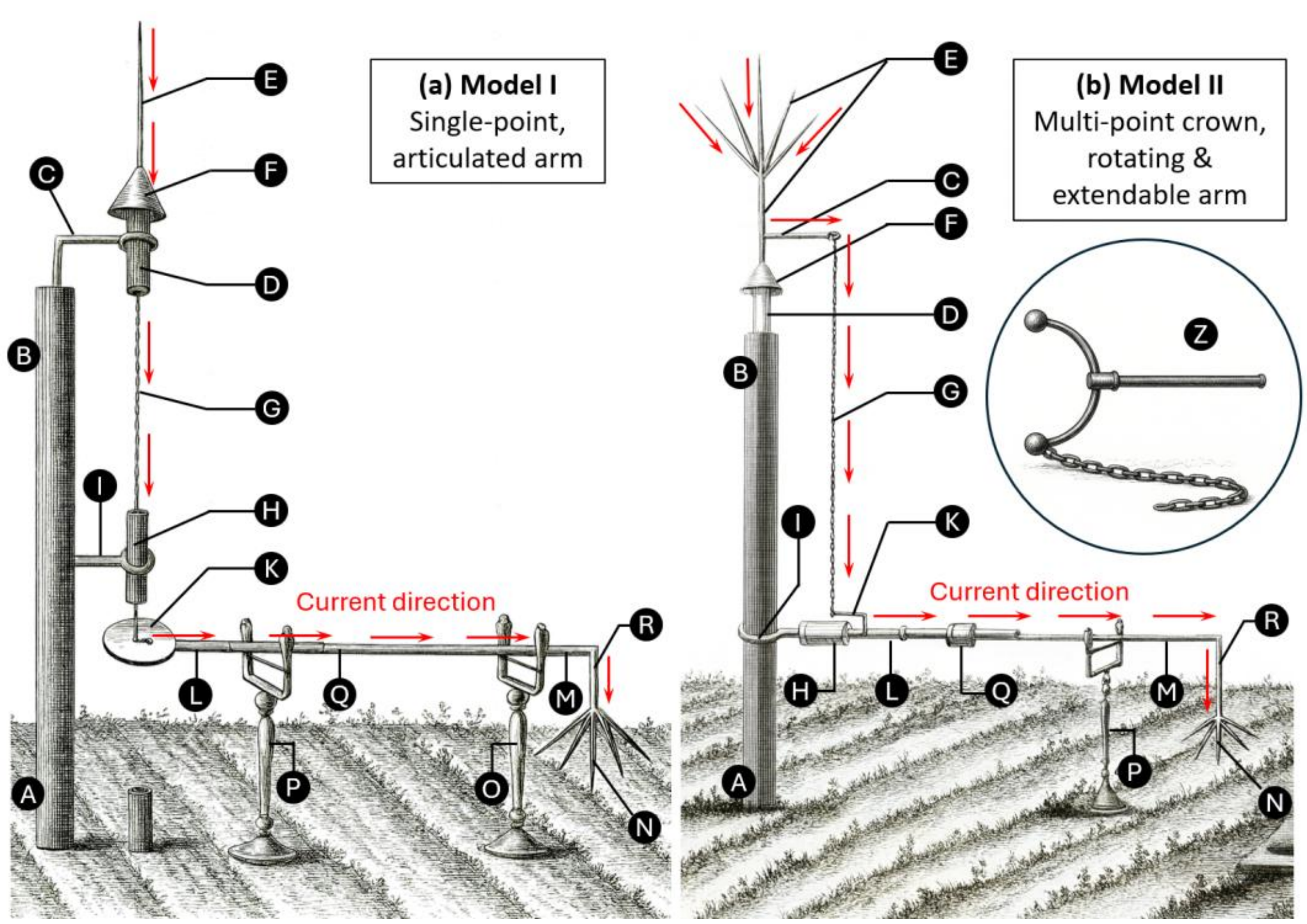

***Figure 6. Bertholon's électro-végétomètre, devised in the 1780s to channel "atmospheric electricity" toward crops. Two variants are shown: (a) Model I, with a single-point collector and articulated arm distributing charge to pointed outlets; and (b) Model II, with a crown of points, a rotating extendable arm and broader reach. Both relied on slow diffusion of charge through sharp points, conceived as a gentle stimulus to vegetation and included a neutralization device (Z) for safety, represented in inset of (b). Adapted from [31].***

The first model (Figure 6a) which may be termed the "single point, articulated arm". It consisted of a wooden mast firmly anchored and protected against moisture. At the top, a single iron point served as a collector. It was shielded from rain and insulated from the electrical ground by a glass tube filled with bituminous mastic. The captured charge descended via a conducting chain to an insulated, articulated horizontal arm that could be aimed over crop rows. The arm ended in a comb of points that release charge gradually as bluish corona glows (aigrettes lumineuses), which Bertholon believed would stimulate vegetation.

The second model (Figure 6b) which may be termed the "multi-point crown, rotating and extendable arm". It kept the same logic while boosting collection and coverage. The mast head received a resin-impregnated insert to improve insulation from the ground and the single point was replaced by a cluster of sharp points to enhance charge extraction. The transmission still used a vertical chain that was connected to the horizontal arm. The arm could rotate 360° and was length-adjustable, allowing one to sweep the entire plot. Again, its tip carried diffusor points.

For clarity, Table 1 provides a comparative summary of the two variants, detailing their structural components, operating principles and distinctive features, in order to highlight both their shared logic and their technical differences. Both models included a "C-shaped exciter": a rudimentary grounding tool designed for operator safety and represented in the inset of Figure 6b. It consisted of a curved metallic rod with rounded ends, one of which carried a chain trailing on the soil to provide an earth connection. The user held the device by an insulated glass handle. When brought near the electro-végétomètre, the metallic arc collected excess atmospheric charge, which was harmlessly discharged through the chain into the ground. In modern terms, it functioned much like a primitive discharge wand still used in high-voltage laboratories to neutralize charged equipment safely.

Bertholon imagined a second "dry-route" method to target wood-boring insects hidden in branches, trunks or roots (Figure 7). He proposed to discharge a Leyden jar directly through the affected wood by positioning two iron wires on opposite sides and releasing the stored charge across the gap. This brief high-voltage pulse would result in a strong electric field forcing a transient current through the tissue, with the aim of killing larvae inside. By shifting the wires and repeating the operation, he imagined one could "sweep" the branch until the pest was eliminated. For orchards, he extended the design into a rudimentary electrical grid: trees were linked with bare iron wires so that a single discharge could propagate through successive spans, exposing multiple trunks at once. He even recommended driving current through soil paths to target roots or applying preventive shocks during the season when the infestation typically began. As with the electro-végétomètre, this was not measurement but mediation in intent: a scheme to channel atmospheric or stored electricity into plant structures. No dosimetry of voltage, current or energy was attempted, but the procedure illustrates Bertholon’s transition from abstract conjecture to technical intervention: building circuits, defining contact points and integrating electrical pulses into common agricultural practices.

***Table 1. Comparison between the design and operation of Bertholon's electro-vegetometer (Fig. 6): variant I ("single-point articulated arm") and variant II ("multi-point crown, rotating and extendable arm"). The letters (A to Z) correspond to the captions of the original figures. Both devices are passive collectors-distributors (no field/charge/current measurement); the terminal diffuser emits a charge in the form of a non-disruptive corona above the crop.***

| | **Model I**<br>**Single-point, articulated arm** | **Model II**<br>**Multi-point crown, rotating & extendable arm** |
|---|---|---|
| **Purpose** | Passive capture → Insulated conduction → Gentle diffusion over crops (no measurement). | |
| **Mast / Base** | Wooden mast (A–B): buried section fire-dried, tarred, wrapped in charcoal dust/cement, then masonry footing; above-ground part painted/bitumen-coated for weather resistance. | Mast (A–B) with resin-saturated head cylinder (C) treated with tar/pitch/turpentine to improve insulation from ground. |
| **Top insulation** | - Thick glass tube packed with bituminous mastic (D)<br>- Second glass insulator downstream (H). | Insulated rotary interface via resin insert (D) and glass/mastic sleeve (H). |
| **Collector (at mast head)** | Single pointed rod (E) mounted in D; protected by a tinplate funnel (F). | Multipoint collector (E) (sharp "crown") to maximize charge capture. |
| **Conduction path (vertical)** | Conducting chain (G) suspended from E, guided through H to prevent leakage. | Bent iron lever (C) carrying chain (G) to the arm interface. |
| **Regulation / buffer** | Chain terminates at iron disk (K) acting as a small "condenser/regulator" to buffer atmospheric variations. | No separate disk; regulation implicit (focus on mobility and reach). |
| **Horizontal arm / distribution** | Articulated conductor arm (K-L-M-N) hinged at L and Q, supported on insulating trestles (O, P) with stretched silk cords. | Telescoping conductor arm (L-M) with its telescoping conductor (Q), supported on insulating trestle (P) with stretched silk cords. |
| **Terminal diffuser** | Comb/cluster of sharp metal points at the arm tip (N, R), releasing charge as non-disruptive corona over the crop canopy/soil. | |
| **Mobility / coverage** | Arm swings on hinges (L, Q) to cover a sector (row/strip). | Full 360° rotation via I-L-M & extend/retract (Q) to sweep a full circle and vary reach. |
| **Neutralization (on/off control)** | Apply a chain (or vertical rod) between K and the ground to break insulation and bleed charge. Also used to null effect when "excess" of electricity is presumed. | |
| **Operation (per Bertholon)** | Use when the air is "poor in electric fluid"; distribute via N; ground at K when "excess" is observed. | |
| **Safety** | Sharp points release corona glows (no sparks). "C-shaped exciter" (copper/iron frame with glass handle + trailing ground chain) recommended to shunt charge during handling. | |

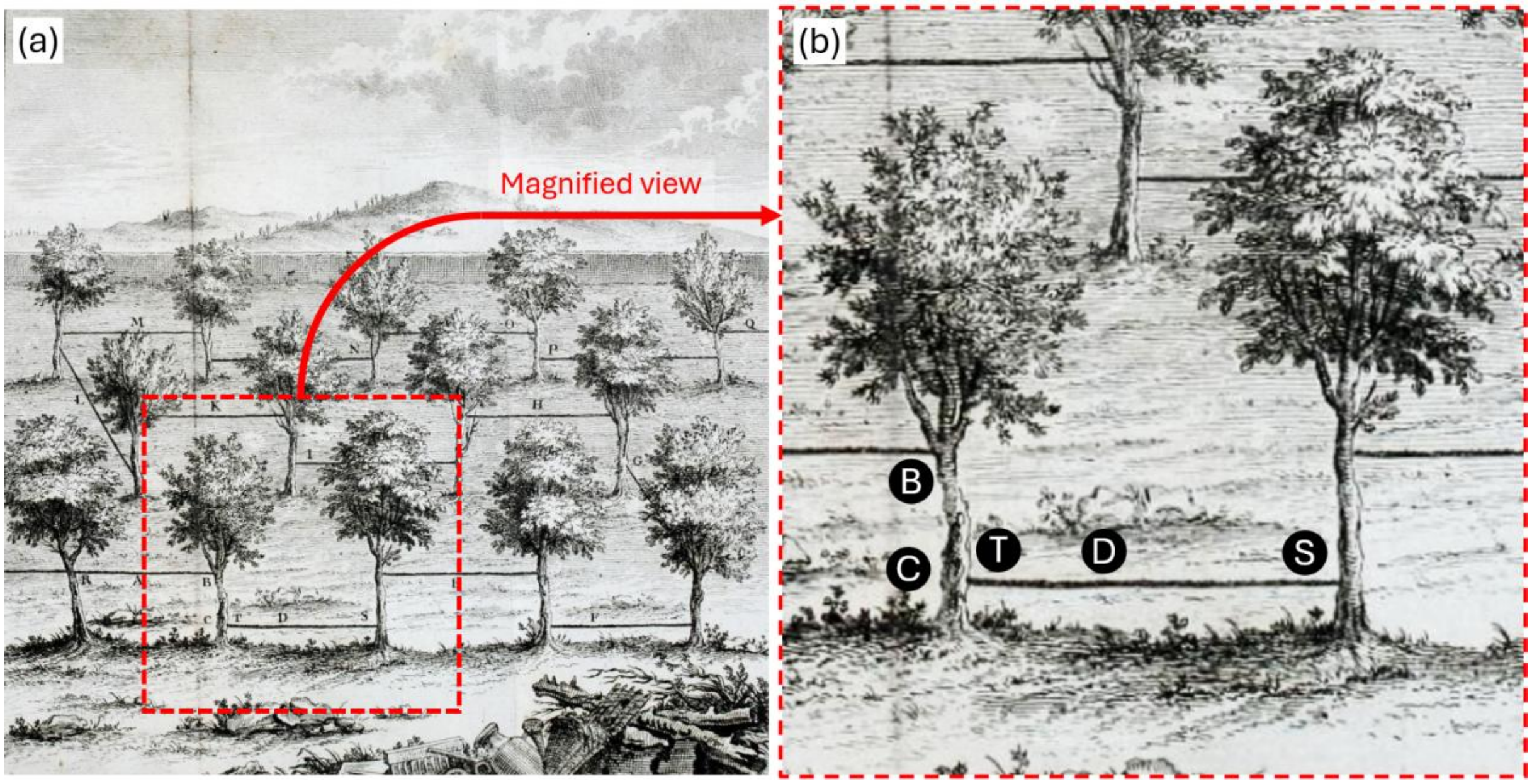

***Figure 7. (a) General view of a fruit orchard where trees are interconnected by uninsulated iron wires forming a rectangular grid (segments A–Q). A Leyden jar (not shown) is discharged across the network, with its outer and inner coatings connected respectively to the first and last wire, so that a single electrical pulse traverses successive spans and penetrates branches, trunks or roots where boring larvae are suspected. Repeated discharges allow the operator to "sweep" the orchard, while roots can be included by inserting opposed wires directly into the soil. (b) Close-up of one grid interval: two wires bracket a trunk section (B–C), so that a Leyden discharge is forced through the wood, targeting larvae hidden in galleries. Additional labeled spans (e.g., T-D-S) illustrate continuity of the circuit to neighboring trees and plant parts. Adapted from [31].***

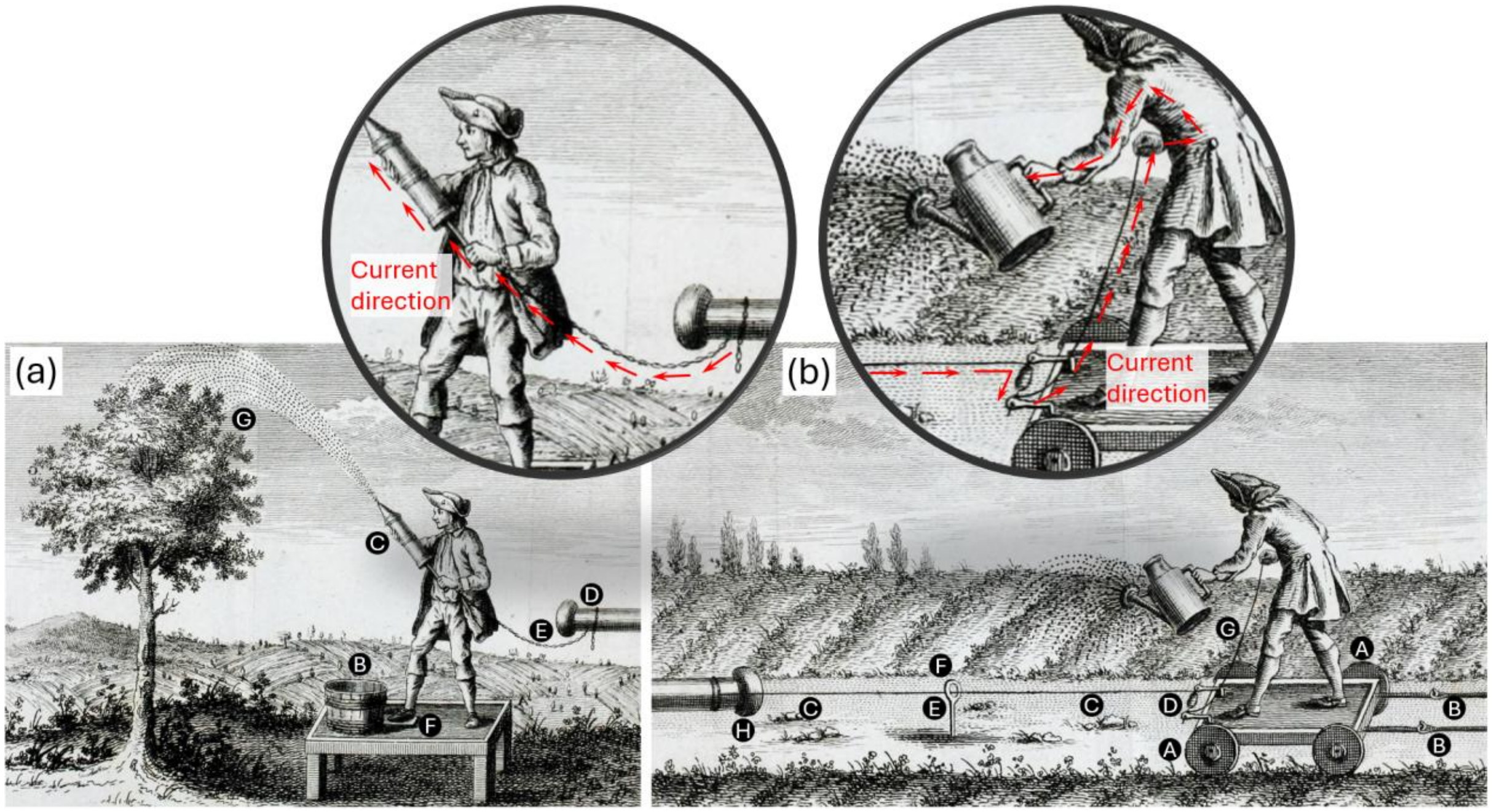

***Figure 8. (Left) "Artificial electrical watering". An electrostatic machine feeds a workstation on an insulating stool with bucket and syringe-pump; the spray is charged before reaching the canopy. The goal is practical (to integrate electricity into routine watering). Insulation keeps body current negligible and directs it with the jet into plants/soil. (Right) Mobile "electric watering" for beds. A resin-insulated cart unwinds a wire from the machine, supported on insulating posts. Holding the live cord, the operator pours from a can, producing "electric rain" along the row. Designed for cadence (assistant refills cans); insulation protects the operator; useful current travels with the water. Adapted from [31].***

#### 2.3.2. Water-route devices: electrified irrigation & mobile systems

Bertholon also devised a second family of instruments based on irrigation with “electrified water”, this time powered by an electrostatic machine rather than relying on the atmosphere. Two main configurations were proposed (Figure 8).

The first, intended for trees (Figure 8a), placed the operator on an insulating stool holding a bucket and syringe pump. The bucket was directly connected to the electrostatic machine, rendering its contents conductive. Each spray of water was thus electrically charged before reaching the canopy. Safety depended entirely on insulation: as long as the operator remained isolated from the ground, the body current was negligible, with visible sparks or tingling confined to the water jet.

The second system was designed for vegetable beds (Figure 8b). Here, an insulating resin cart carried a reel of wire connected to the machine. As the cart advanced along the rows, the operator, holding the live cord, poured from watering cans so that the falling stream produced an “electric rain”. The workflow was organized for efficiency: a second gardener refilled cans continuously while the cart advanced. Again, insulation ensured that the useful current traveled with the water into the plants and soil, rather than through the operator.

Both devices were described as inexpensive, adaptable and effective when applied repeatedly over several days to seeds or growing plants. Their operating principle was not metrological but practical: to embed the action of electricity into everyday irrigation routines. The comparative features of these two systems are summarized in Table 2, which contrasts their design, workflow, safety provisions and intended agronomic applications.

***Table 2. Comparison of Bertholon's devices for electrifying irrigation water (Fig. 8). Tree-by-tree irrigation (Fig. 8a) used an insulating stool, bucket and syringe-pump, while the mobile cart system (Fig. 8b) adapted the principle to vegetable beds, with a resin cart unwinding wire and electrifying watering cans. Both designs aimed to embed electricity into routine watering practices; safety relied on insulation from ground, with useful current carried by the water jet into soil and plants.***

| Feature | Tree-by-tree irrigation | Mobile cart for vegetable beds |
|---|---|---|
| **Insulation of operator** | Operator stands on insulating stool (A), a board supported by glass pillars. | Operator stands on resin-molded insulating cart (A) drawn along garden rows. |
| **Water source** | Bucket (B) placed on the stool platform. | Watering cans refilled continuously by a second gardener. |
| **Irrigation device** | Syringe-pump (C) sprays water onto tree canopy (G). | Operator pours from watering can while holding live cord (G), producing “electric rain” over crops. |
| **Electrical connection** | Electrostatic machine connected by chain (E) from conductor (D) to workstation. | Cart carries reel (D) with insulated axle, unwinding wire (C) connected to machine conductor (H). |
| **Conduction to water** | Tinplate plate (F) under operator's foot ensures bucket communicates with charge. | Wire passes through insulating supports (E) with glass rings (F); additional supports can be added. |
| **Mode of electrification** | Water in bucket becomes charged; sparks and tingling visible at jet or on a metal plate. | Watering can and operator at same potential; electrified water falls as “electric rain” onto crops. |
| **Mobility / workflow** | Device carried tree to tree; simple and inexpensive. | Cart advances row by row; workflow optimized with continuous supply of full watering cans. |
| **Safety** | Insulation of stool ensures current flows into water rather than operator. Minor tingling only if grounding path provided. | Operator insulated on resin cart; negligible current through body unless contact with earth occurs. “C-shaped exciter” recommended for safe approach/withdrawal. |
| **Agronomic use** | Claimed effective when repeated daily over seedlings or growing plants. | Described as slightly slower than ordinary irrigation but with clear agricultural benefits when repeated over several days. |

## 2.4. Technical and methodological limitations

By the late eighteenth century, Bertholon's program was limited above all by instrumentation. Lane's spark electrometer (1767) and Henley's quadrant (c. 1770) provided only coarse, qualitative indicators (spark counts or angular deflections) that could at best approximate the charge stored in a Leyden jar **[72,73]**. They offered no means to quantify the actual energy delivered to plant tissues or the flux of charge at a biological target. Coulomb's torsion balance (1785), though it established the first quantitative law of electrostatics, post-dated Bertholon's writings and measured forces rather than energy deposition, leaving no path toward calibrated protocols for agricultural electrification **[74, 75]**.

The conceptual framework was equally incomplete. Several major agents of electrical discharges remained unidentified: ozone, although its pungent odor was remarked by van Marum in 1785, was not recognized as a distinct chemical species until 1840 **[76]**, while ultraviolet radiation was discovered only in 1801 **[77, 78]**. Reactive oxygen and nitrogen species (RONS), as well as electrohydrodynamic flows, were entirely unknown. In this context, the functioning of Bertholon's electro-végétomètre raises important retrospective questions: was its terminal diffuser (a comb or cluster of sharp metal points) capable of generating fields strong enough to act on vegetation? Could it have produced a form of cold plasma, even if the term was not yet available? In *De l'Électricité des Végétaux* (pp. 399, 401), Bertholon himself reported luminous and electrical aigrettes visible at night on the multi-point terminals of his device. Modern physics recognizes these bluish-violet glows as corona discharges: localized plasmas produced when strong fields at sharp tips accelerate electrons. Such discharges release charge gradually, without sparks and generate species unknown in Bertholon's time: ions, electrons, ozone and nitrogen oxides [31, 79]. To Bertholon, however, they appeared as visible manifestations of the "electric fluid" nourishing plants.

Experimental practice, too, fell well short of modern standards. Bertholon relied on qualitative accounts of germination and growth, with no systematic replication, randomization or formal control groups. Trials were conducted in open air and thus remained fully exposed to meteorological variability, including humidity, temperature, soil conductivity, rainfall and illumination. Even the instruments available to him, such as Henley's or de Saussure's electrometers, yielded only relative indices rather than traceable measurements. The methodological tools required to stabilize inference (i.e. randomized designs, comparative statistics and variance analysis) would only appear in the 20th century [80, 81].

Finally, the electrostatic machines themselves lacked regulation or stability. Their output depended on crank speed, ambient humidity and electrode condition. Accessories such as Leyden jars or pointed conductors could channel or limit discharges but did not measure them. Lane's discharger made it possible to fix a spark gap, yet it provided no information about the actual dose delivered to plants. As a result, pulse energy, repetition rate and source–plant geometry remained uncontrolled, precluding the establishment of reproducible dose–response curves.

In contrast, modern plasma agriculture defines "dose" through explicit, quantitative parameters: electric field strength integrated over time ($V \cdot m^{-1} \cdot s$), surface charge density ($C \cdot m^{-2}$) and specific or areal energy deposition ($J \cdot kg^{-1}$ or $J \cdot m^{-2}$), all referenced in carefully controlled source–target geometries [54, 56, 82]. Today, the respective contributions of fields, RONS, photons, localized heating and electrohydrodynamic flows can be separated and quantified, enabling reproducible mechanistic studies that were unattainable in Bertholon's time.

# 3. Modern physics of cold plasmas applied to agriculture

While Bertholon framed plant electrification through the metaphor of a subtle "electric fluid," modern science grounds the discussion in plasma physics. The shift from qualitative impressions (such as visible aigrettes on metallic points) to quantitative descriptors of charge, fields and species requires defining what a plasma is, the parameters governing its behavior and the diagnostic methods that render it measurable.

## 3.1. Definition of cold plasma & Key properties

Plasma is commonly known as the fourth state of matter, distinct from solids, liquids and gases. It is characterized by macroscopic quasi-neutrality (i.e., approximately balanced positive-ion and electron densities, as proposed in Equation (1)) and by collective dynamics dominated by long-range Coulomb interactions. Two fundamental parameters govern its electrostatic behavior: the Debye length $\lambda_D$ (Equation (2)) and the electron plasma frequency $\omega_p$ (Equation (3)), which depend on the vacuum permittivity $\varepsilon_0$, the Boltzmann constant $k_B$, the elementary charge e, the electron temperature $T_e$ and of the electron density ne [83]. A medium qualifies as a plasma when the Debye length is much smaller than the system size L (Equation (4)), the number of particles in a Debye sphere $N_D$ is $\gg$1 (Equation (5)) and collective effects dominate over binary collisions (Equation (6)) [84].

$$n_i \approx n_e \quad \{1\}$$

$$\lambda_D = \sqrt{\frac{\varepsilon_0 k_B T_e}{n_e e^2}} \quad \{2\}$$

$$\omega_p = \sqrt{\frac{n_e e^2}{\varepsilon_0 m_e}} \quad \{3\}$$

$$\lambda_D \ll L \quad \{4\}$$

$$N_D = \frac{4}{3}\pi n_e \lambda_D^3 \quad \{5\}$$

$$N_D \gg 1 \quad \{6\}$$

Depending on density and temperature, plasmas range from strongly collisional (when the mean free path is comparable to or smaller than the system size) to weakly collisional, as in magnetospheric plasmas or low-pressure RF discharges. In air at terrestrial pressure, electrons of a few eV have a mean free path of approximately 0.1–1 µm. Near boundaries, quasi-neutrality breaks down within space-charge sheaths whose thickness is a few $\lambda_D$; these sheaths control how current, energy and chemistry couple to surfaces [84, 85].

A cold atmospheric plasma (CAP; also non-thermal plasma, NTP) is a non-equilibrium discharge in which the electron temperature $T_e$ (≈1–5 eV, i.e., 11 600–58 000 K) greatly exceeds the gas temperature $T_g$ (≈300–350 K). At atmospheric pressure, strong collisionality leads to regimes dominated by electron–molecule and ion–molecule collisions and to thin sheaths at electrode–gas or gas–solid interfaces. Depending on configuration and excitation, one observes streamer regimes (channeled micro-discharges <10 ns, reduced field E/N ≈100–300 Td) and dielectric-barrier discharges (DBDs), which may be filamentary (E/N ≈150–200 Td) or diffuse (<100 Td) [84]. The reduced field E/N (1 Td = $10^{-21}$ $V \cdot m^2$) shapes the electron energy distribution function (typically non-Maxwellian) and thus the probabilities of excitation, ionization and dissociation through energy-dependent cross sections, directly influencing overall reaction kinetics [86].

CAPs couple multiple processes. The electric field drives charge transport and interface polarization while the chemical component is carried by reactive oxygen and nitrogen species (RONS). Those include short-lived (e.g., $O(^3P)$ µs–ms, $OH^\bullet$ ≈ µs, $^1O_2$ ≈ ms, $NO^\bullet$ ≈ µs) and long-lived (e.g., $O_3$ s–min, $H_2O_2$ s–h, $NO_2^-/NO_3^-$ stable in aqueous phase) [87, 88]. These species are produced and transported together with UV/visible emission from molecular bands (e.g., $N_2$ second positive system, $OH^*$) and with electrochemical gradients at plasma–gas, plasma–solid and plasma–liquid interfaces. The energy chain can be summarized: source → electrons → elementary reactions → reactive species → interfacial fluxes [89].

Common atmospheric sources include DBDs (planar, coaxial or reactor geometries) driven sinusoidally or in pulsed mode (kHz–MHz; ns–µs pulses) [90] and atmospheric pressure plasma jets (APPJ) operating in helium or argon with small $O_2/H_2O$ admixtures [91]. Control parameters (gap and geometry, dielectric material and thickness, waveform, amplitude, frequency, duty cycle, gas composition and flow, relative humidity) set E/N, the specific energy deposition, discharge morphology (filamentary vs. diffuse, streamer density) and the spectrum of species produced.

Metrological qualification is essential for reproducibility and inter-comparison. Electrical diagnostics include recording V (t) and I(t), estimating average power via Lissajous figures and evaluating energy per pulse [92]. Temporal dynamics are resolved with ICCD imaging (ns–µs gates) to visualize streamer propagation and discharge morphology [93]. Gas temperature can be inferred from rotational spectroscopy of $N_2$ or OH bands and surface temperature by infrared thermography [94–96]. Physico-chemical analysis uses optical emission spectroscopy to identify excited species and, with kinetic modeling, to estimate effective electron energies. In the aqueous phase, routine tracking includes $H_2O_2$, $NO_2^-$, $NO_3^-$, pH and conductivity, with attention to their temporal stability [87].

Across applications, good practice entails arc suppression, monitoring sheath evolution and controlling surface heating. Reporting conditions in terms of specific energy (e.g., $J{\cdot}cm^{-3}$ of treated gas or activated liquid) and/or equivalent chemical indicators enables robust links between process, dose and response, independent of the particular agricultural use case.

## 3.2. Technologies and laboratory protocols

In agronomy, cold atmospheric plasma (CAP) is implemented through three complementary routes that target different biological compartments: seeds, seedlings, leaves/canopies, whole plants, fruits, soils and waters. These are:

- DDS (Direct Dry-Surface exposure): plasma applied directly to tissues with no intentional free water at the interface.
- DWS (Direct Wet-Surface exposure): plasma applied in the presence of a thin aqueous film at the interface.
- PAW (Plasma-Activated Water): water or nutrient solution treated ex situ by plasma, then applied via soaking, spraying, irrigation/fertigation, hydroponics or postharvest rinsing [58].

In practice, laboratories mainly rely on DBD and APPJ for DDS/DWS. DBD panels or meshes are suitable for trays, awnings and fruit lines; APPJs (He or Ar with traces of $O_2/H_2O$) offer directional diffusion to localized tissues (wounds, grafts) and hard-to-reach geometries. Corona arrays not only release charge but also generate a small electrohydrodynamic (EHD) airflow, which disturbs the still boundary layer of air on leaf surfaces and thereby improves the delivery of plasma-generated species to plant tissues. For PAW generation and process-water treatment, sliding arcs and DBD/pin-to-liquid jets are commonly employed [90, 91, 97]. The indicative operating ranges include:

- DBD: 6–20 $kV_{pp}$, 5–50 kHz, gap 0.5–5 mm, exposure 0.5–60 s.
- APPJ: 2–10 $kV_{pp}$, 10–200 kHz or RF, 0.5–5 $L{\cdot}min^{-1}$ He/Ar with traces of $O_2/H_2O$, nozzle-to-tissue distance of 5–30 mm.

The dose is expressed in units that are independent of the device and related to the exposed entity:

- For direct exposures (DDS/DWS), it can be expressed in specific energy relative to the biological load, for example $J{\cdot}m^{-2}$ for leaves and $J{\cdot}kg^{-1}$ for seed lots, as well as electrical operating conditions (waveform, pulse width and repetition frequency, peak-to-peak voltage, energy per pulse and average power), gas composition/flow rate, source-target distance, scan speed, exposure time and effluent temperature. Where applicable, quantification of UV fluence at the tissue level and emitted gas concentrations ($O_3$, $NO_2$/NO) can further clarify the regime [98, 99].
- For PAW, characterization generally includes the specific electrical energy transmitted to the liquid ($J{\cdot}L^{-1}$), the chemical state at the time of application ($H_2O_2$, $NO_2^-/NO_3^-$, pH, conductivity) and the dose applied ($mL{\cdot}m^{-2}$ or per plant, with contact time) [87, 100, 101]. Where possible, a surface-normalized marker load (e.g., µmol $H_2O_2$ $m^{-2}$) is informative.
- In all cases, noting the time between activation and use ($t_0$, $t_0 + \Delta t$ ) and reporting measurement uncertainties facilitates comparison between devices and laboratories.

## 3.3. Effects of cold plasma on seeds

CAP exerts three major, interdependent influences on seeds. First, they alter the physico-chemistry of the seed coat, increasing surface hydrophilicity and effective permeability. Second, they reduce (sometimes nearly eliminate) microbial load and seed-borne pathogens. Third, they trigger early physiological and molecular responses, including hormonal adjustments, activation of antioxidant defenses and transcriptional reprogramming. These processes are interlinked and together help explain the consistent improvements in germination and early growth reported across species. The following sections detail each of these three aspects in turn.

#### 3.3.1. Seed-coat modification and wettability.

Microscopy studies (AFM/SEM) consistently show that cold atmospheric plasma (CAP) alters the outer surface of seeds by roughening and restructuring the coat through processes such as etching, ablation and nano/micro-structuring. These changes directly improve the ability of the seed to absorb water. In wheat and barley, for instance, Figure 9 shows that $Ar–O_2$ plasma treatments produced visibly fissured, rougher surfaces that correlated with increased water uptake, though the magnitude of this effect depended on the treatment protocol [102, 103]. Glow discharge regimes produced strong etching and nanostructuring, whereas afterglow exposures had little measurable effect [104]; extending the exposure time further enhanced surface roughness [105]. Comparable modifications have been observed in maize exposed to Ar–$O_2$ low-frequency glow discharges [106]. Surface chemistry is also transformed. Analyses using XPS and FTIR have shown that plasma introduces oxygen-containing functional groups on the seed surface, indicating oxidation. This chemical modification increases surface energy and is stable over time, with no hydrophobic recovery after treatment [107, 108]. One of the clearest experimental signatures of these changes is the dramatic reduction in water contact angle (WCA). In lentil, bean and wheat, RF air plasma reduced WCA values from 127°, 98° and 115° down to 20°, 53° and nearly 0° (complete wetting), respectively (Figure 10) [109]. This sustained increase in hydrophilicity accelerates imbibition, the crucial first step in breaking dormancy and promoting uniform germination [108, 110].

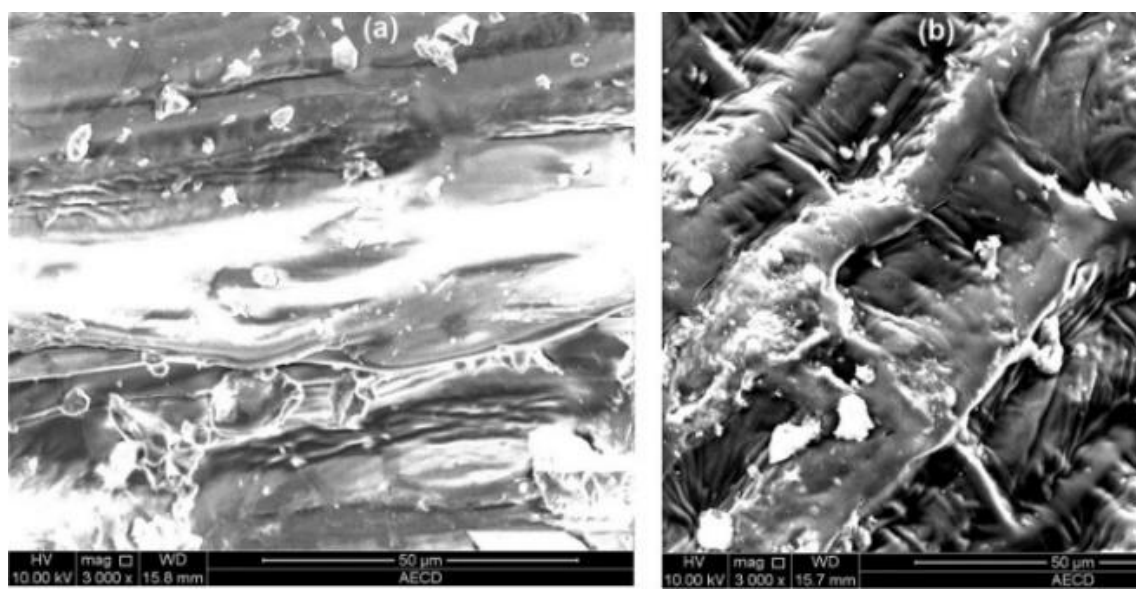

*Figure 9. SEM images of wheat seed surfaces treated for 90 s with (a) control (no plasma), (b) Ar/$O_2$ plasma. Scale bar is 50 µm. Reproduced from M.M Rahman et al. Scientific Reports, Springer Nature, 2018, licensed under CC BY 4.0 [102].*

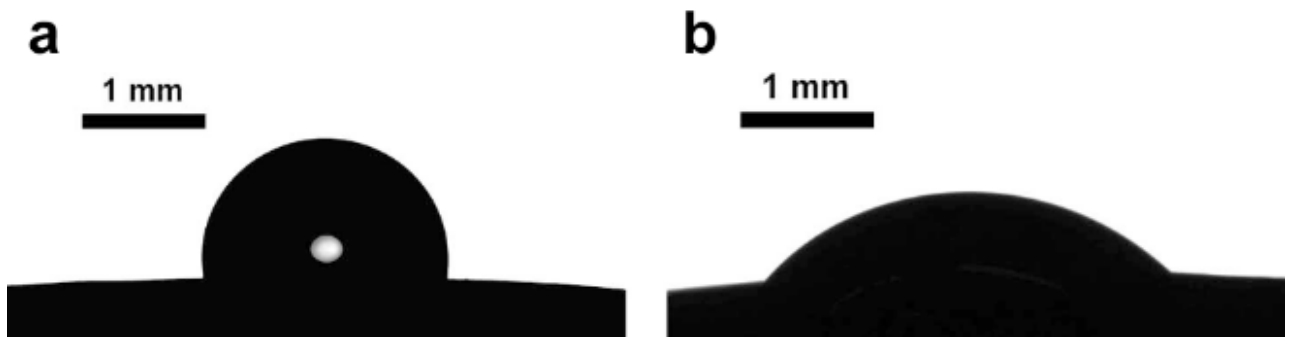

*Figure 10. Water droplet deposited on a bean seed which is (a) untreated or (b) plasma-treated. Reproduced from Bormashenko et al. Scientific Reports, Springer Nature, 2012, licensed under CC BY 4.0 [109].*

#### 3.3.2. Decontamination and pathogen control.

CAP treatments frequently achieve substantial reductions in surface microbial loads for a wide variety of seeds (maize, wheat, sunflower, …) [111]. For instance, Figure 11a demonstrates effective fungal decontamination of sunflower seeds following a 5-min plasma treatment achieved in ambient air. Similarly, Figure 11b indicates that 2–5 minutes of CAP exposure in chickpea (Cicer arietinum) lowered natural microbiota by 1–2 log units, with a 1-minute treatment proving optimal by both maximizing germination (89.2%) and reducing median germination time to 2.7 days [112]. On hazelnuts, low-pressure $SF_6$ plasma achieved up to a 5-log reduction of Aspergillus parasiticus, as highlighted in Figure 11c [113]. However, complete microbial inactivation remains uncommon and strongly species-dependent. In sweet basil, DBD treatments applied for 10–15 minutes significantly reduced Fusarium oxysporum f. sp. basilici colony-forming units in a clear dose–response pattern, yet viable colonies persisted within seed recesses, highlighting the limited penetration of RONS and the resilience of certain propagules (e.g., chlamydospores). Extended treatment durations further decreased germination capacity, particularly after 6 months of storage, underscoring a trade-off between effective decontamination and seed viability [114]. At the community level, taxa such as Alternaria and Epicoccum show comparatively high tolerance and in buckwheat species, exposure beyond 90–120 s led to sharp declines in germination, rendering treatments unsuitable for sowing when doses are miscalibrated [115].

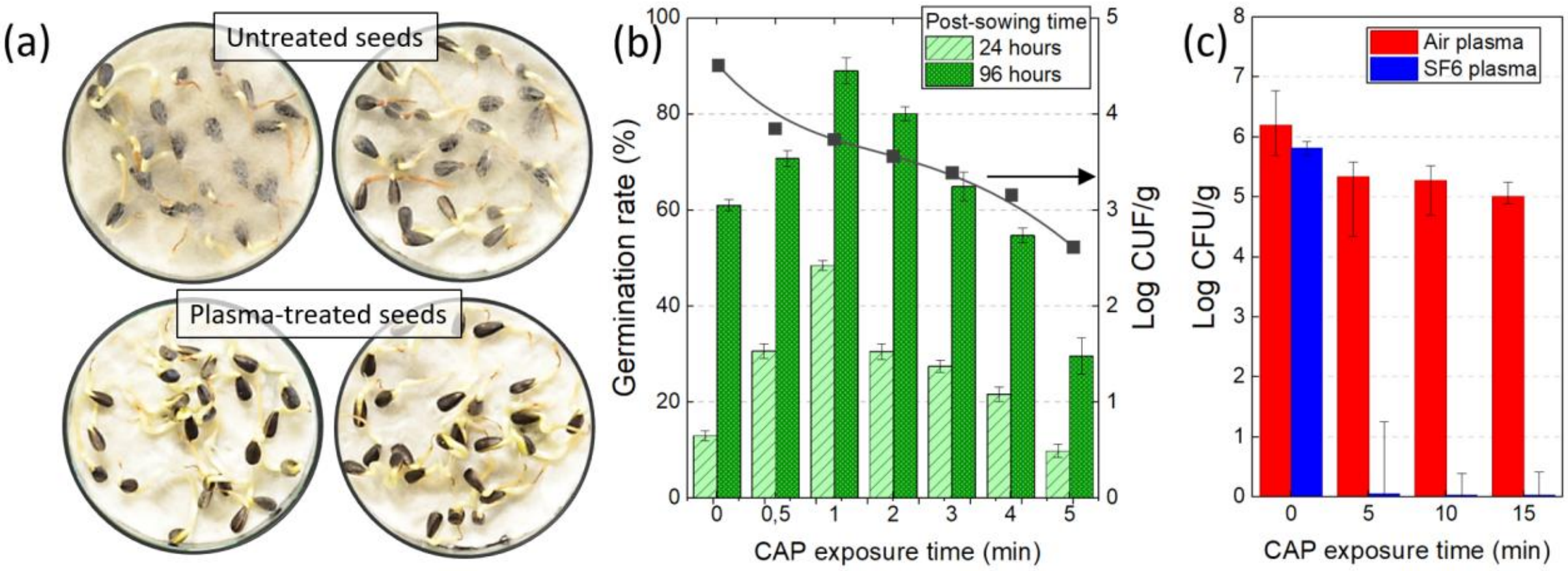

*Figure 11. (a) Photographs of sunflower seeds contaminated by Rhizopus either untreated or plasma-treated for 5 min (T. Dufour, LPP). (b) Influence of CAP treatment on the germination of Cicer arietinum seeds (left axis) and on the survival of surface micro-organisms (right axis). The bars show mean value +/- SEM germination (%) (n = 6; $p<0.05$) Adapted from [112] (c) Reduction effect of air plasma and $SF_6$ plasma on A. parasiticus contaminated hazelnuts for 15 min inoculation. Adapted from [113].*

#### 3.3.3. Physiological and molecular responses.

Mechanistically, plasma-generated RONS promote abscisic acid (ABA) catabolism, enhance gibberellic acid (GA) signaling, interact with nitric oxide (NO) and induce the expression of cell wall–weakening genes in the endosperm and testa, thereby lowering barriers to dormancy release. In Arabidopsis, Grainge et al., demonstrated that PAW breaks dormancy through combined $NO_3^-$, $H_2O_2$, $NO^\bullet$ and $OH^\bullet$ signals, with distinct responses observed in ABA/GA pathway mutants (e.g., cyp707a2, nlp8, prt6), alongside activation of cell wall remodeling genes [116]. A companion study by the same group showed that PAW also modulates very low- and low-fluence phytochrome responses (VLFR/LFR; PHYA/PHYB), thereby extending benefits to photodorsmant seeds [117].

At the physiological level, one of the most consistent outcomes is the up-regulation of antioxidant defenses. In oilseed rape seeds exposed to drought, CAP improved germination parameters and also enhanced post-germinative physiological responses. As reported in Figure 12, the superoxide dismutase (SOD) and catalase (CAT) activities were enhanced while lipid peroxidation (MDA) was reduced and the accumulation of osmoprotectants was enhanced [118]. In rice under cold stress, argon DBD accelerated germination and seedling growth, elevated SOD/CAT/peroxidase (POD) activity and triggered transcriptomic shifts involving thousands of differentially expressed genes, including suppression of NCED3 (ABA biosynthesis) and induction of stress-tolerance genes [119]. Similar trends were reported in soybean, where CAP induced dose-dependent increases in SOD, CAT and ascorbate peroxidase (APX), improving both germination indices and vigor [120].

Overall, these effects often extend beyond germination. In wheat, PAW enhanced early establishment by increasing photosynthetic pigments and soluble protein content, suggesting a metabolic “pre-priming” effect beyond simple microbial decontamination [121]. In maize, gliding-arc exposure modified hormone pools while simultaneously stimulating seedling establishment [122].

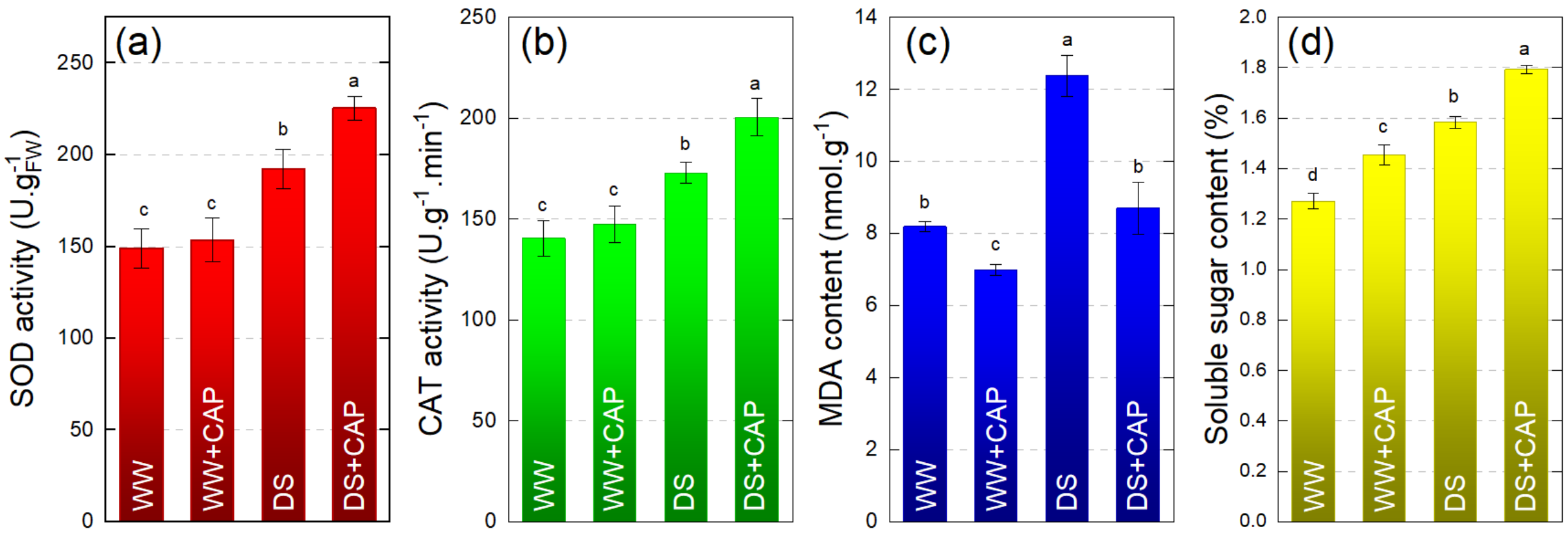

***Figure 12. Effect of drought stress and cold plasma treatment on 4 biomarkers of physiological state and stress response in the case of oilseed rape seedlings. (a) SOD activity, (b) CAT activity, (c) MDA contents, (d) soluble sugar contents. In all subfigures: WW: well watered; WW + CAP: well watered + plasma; DS: drought stress; DS + Plasma: drought stress + plasma. Error bars indicate standard error (n = 3). Different lowercase letters denote statistical differences between treatment groups at the 5% level according to Duncan’s test. Adapted from [118].***

## 3.4. Effects of cold plasma on seedlings and plants

CAP (whether through DDS, DWS or PAW) can stimulate growth and stress tolerance in multiple plant species, though outcomes remain strongly dose- and context-dependent. Across species, CAP or PAW can increase seedling biomass, leaf area and chlorophyll indices. For example, PAW promoted tomato seedling growth and nutritional quality (polyphenols, lycopene), while basil exposed to DBD showed elevated chlorophylls, carotenoids and antioxidant activity with reduced ionic leakage [123, 124]. Long-term benefits of PAW irrigation are visible in tomato and pepper, where seedlings treated with PAW (with or without seed priming) remained larger and more vigorous even 60 days after sowing (Figure 13a for tomato plants, Figure 13b for pepper plants) [125]. In tomato, growth stimulation was also shown to depend on exposure time: moderate PAW treatments (15–30 min) maximized seedling development at 28 days, while longer exposures became less effective (Figure 13c) [123]. Similarly, lentil seedlings irrigated with plasma-activated water (both tap and demineralized) displayed stronger growth compared with controls at 18 days (Figure 13d) [126].

Plasma also affects photosynthetic performance, particularly photosystem II (PSII). Low doses improve photochemical parameters ($\Phi E_0$, $ET_0/RC$), while higher doses inhibit them. For example, proso millet exposed to low discharge power showed enhanced electron transport, but stronger discharges suppressed it [127]. Repeated PAW irrigation in barley improved chlorophyll content and quantum yield, suggesting a potential for longer-term acclimation [128].

Beyond growth promotion, CAP/PAW often upregulates antioxidant enzymes (SOD, CAT, POX) and modulates secondary metabolism. In basil, plasma increased antioxidant activity and peroxidase while stabilizing membranes [124]. In tomato, PAW stimulated salicylic and jasmonic acid pathways and induced pathogenesis-related (PR) gene expression [129]. These responses highlight plasma's role as a defense primer: tomato seedlings sprayed with PAW showed up to 61% reduced disease severity against Xanthomonas, despite no direct antibacterial effect *in vitro*, clearly indicating host-defense activation [130]. Similar priming effects have been reported against pests, e.g., increased spider mite mortality in PAW-treated tomato [131].

The redox-active chemistry of PAW ($H_2O_2$, $NO_2^-$/$NO_3^-$, mild acidification) has also been linked to improved tolerance to salinity and drought. In hydroponic bok choy, plasma-treated nutrient solutions enhanced growth and stress markers, with intracellular RONS implicated [132]. In barley, PAW irrigation supported photosynthetic performance and antioxidant adjustments under saline stress [128]. At the root level, maize showed accelerated endodermal lignification and finer branching under arsenic exposure, suggesting plasma-induced anatomical "fortification" that may improve water and ion uptake [133].

Although the body of evidence supports plasma as a growth and stress-priming tool, responses are highly species- and dose-specific. Beneficial outcomes are typically observed at low exposures, while excessive treatment can impair photosynthesis and growth. Moreover, most studies remain limited to controlled environments, with little information on field robustness, reproducibility across cultivars or long-term yield effects. Clarifying these boundaries is essential before practical deployment in agriculture.

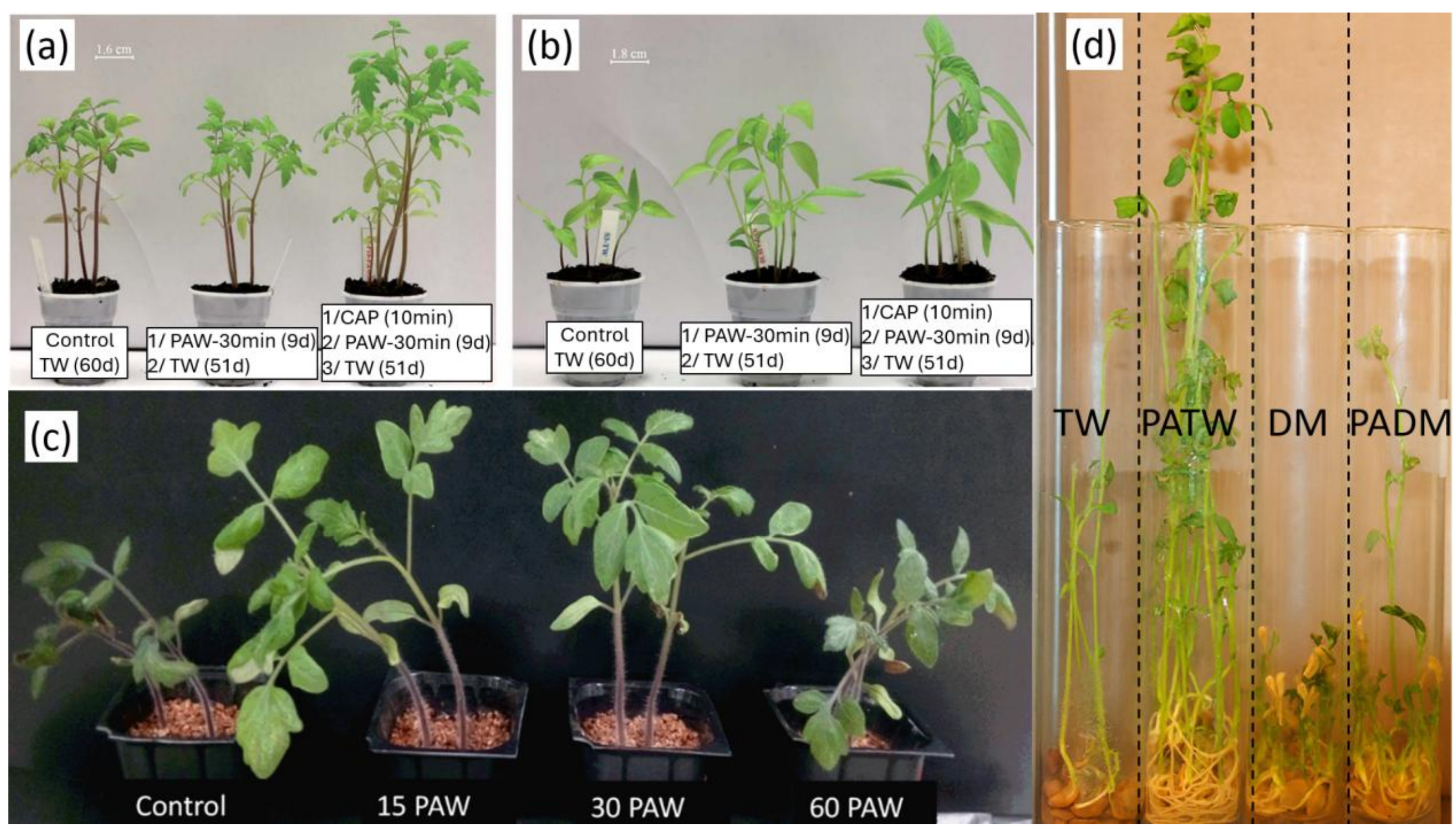

***Figure 13. (a-b) Long-term effect of PAW on tomato seedlings with/without seed priming, observed on 60th day after sowing. (b) Same protocol as (a) for pepper plants. Reproduced from L. Sivachandiran et al, RSC Adv. (2017. Licensed under CC BY 3.0 [125]. (c) Effect of PAW on the growth of tomato seedlings for different exposure times: 0, 15, 30 and 60 min (28 days old). Reproduced from Adhikari et al. Scientific Reports, Springer Nature, licensed under CC BY [123]. (d) Lentil seedlings irrigated either with tap water (TW), plasma-activated tap water (PATW), demineralized water (DM) or plasma-activated demineralized water (PADM). Observation on 18th day. Reproduced from S. Zhang et al., RSC Adv. (2017). Licensed under a Creative Commons Attribution-NonCommercial 3.0 Unported Licence [126].***

# 4. Conclusion & Outlooks

This article has traced a continuous, if discontinuous, line from 18th-century "electrizations" to today's cold-plasma agronomy. The pivot is metrology. Where Bertholon engineered pathways for an indivisible "electric fluid" (capture → conduction → distribution) without access to dose or agents, contemporary plasma agriculture defines sources, quantifies interfacial transfers and measures dose under controlled conditions. Tunable atmospheric DBDs and plasma jets, together with electrical (V(t)/I(t), Lissajous/Manley) and optical diagnostics (OES/absorption), have transformed demonstration into experiment: dose–response curves can be mapped, efficacy windows identified and results replicated across laboratories.

Reappraised in this light, Bertholon's contribution is methodological rather than evidentiary: he supplied an operative vocabulary and the intuition that moderate electrical regimes might beneficially modulate plant functions. Modern work explains when and why that intuition holds. Cold plasmas co-generate fields, RONS, UV radiation, mild heating and electrohydrodynamic flows; diagnostics separate and tune these components. At seed and plant interfaces this yields a coherent mechanistic picture: controlled redox priming and surface conditioning (hydrophilization, micro-etching) accelerate imbibition and germination, reduce surface bioburden, elevate antioxidant capacity and, within defined windows, improve photosynthetic function and stress tolerance. The same platform extends to media (irrigation water/PAW) and the agri-food chain (product and equipment hygiene), provided chemistry and exposure are tracked.

The rupture with historical electroculture is therefore threefold: the nature of the electrical object (from a unitary essence to a multi-physics process), the means of measurement and control (from qualitative signs to chemical/energetic dosimetry) and the standards of validation (from compelling trials to randomized, replicated designs with device-independent reporting). These differences do not validate 18th-century claims; they render them intelligible and, where appropriate, transferable.

Looking ahead, four priorities define an applied agenda. First, anchor practice in traceable dosing for both CAP and PAW routes, reporting source parameters, environmental conditions, seed-lot water status and RONS markers. Second, scale up with eco-designed, air-fed sources and low-energy PAW reactors, evaluating energy cost per agronomic gain. Third, integrate plasma processes within agroecological itineraries, sequencing disinfection with inoculation, pairing with coatings and steering dose by field-relevant indicators. Fourth, ensure safe and robust adoption through ozone/$NO_x$ and UV limits, arc protection, effluent management and multicenter trials across species, varieties and sites.

Seen across three centuries, the motif of an "électricité vivifiante" endures, but its status is transformed. It is no longer a regulative idea grounded in atmospheric reservoirs; it is a mechanistic toolbox (controlled redox signaling, interface engineering, gentle sanitation) capable of delivering lower-input, traceable and potentially more resilient crop systems. The historical bridge matters not as ornament, but as orientation: it reminds us that a sound intuition becomes technology only when it is instrumented, dosed and shared.